\documentclass[useAMS,usenatbib]{mn2e}

\usepackage{graphicx}
\usepackage{mathrsfs}
\usepackage{natbib}
\usepackage{overpic}
\usepackage{xcolor}
\usepackage{amssymb}
\bibpunct[,]{(}{)}{;}{a}{}{,}

\def\totd{{\mathrm{d}}}
\def\sun{{\odot}}

\title[SASI- and Convection-Dominated CCSNe in 3D]{Three-Dimensional
Simulations of SASI- and Convection-Dominated Core-Collapse Supernovae}
\author[Rodrigo Fern\'andez]{Rodrigo Fern\'andez$^{1,2}$\\ \\
$^1$ Department of Physics, University of California, Berkeley, CA 94720, USA\\
$^2$ Department of Astronomy \& Theoretical Astrophysics Center, 
University of California, Berkeley, CA 94720, USA}

\begin{document}

\date{Submitted to MNRAS}

\pagerange{\pageref{firstpage}--\pageref{lastpage}} \pubyear{2012}

\maketitle

\label{firstpage}

\begin{abstract}
We investigate the effect of dimensionality on 
the transition to explosion in neutrino-driven core-collapse 
supernovae. Using parameterized hydrodynamic simulations of the stalled 
supernova shock in one-, two- (2D), and three spatial dimensions (3D), 
we systematically probe the extent to which hydrodynamic instabilities alone 
can tip the balance in favor of explosion. In particular, we focus on
systems that are well into the regimes where the Standing Accretion Shock
Instability (SASI) or neutrino-driven convection dominate the
dynamics, and characterize the difference between them. 
We find that SASI-dominated models can explode with up to
$\sim 20\%$ lower neutrino luminosity in 3D than in 2D, with 
the magnitude of this difference decreasing with increasing resolution.
This improvement in explosion conditions is related to
the ability of
spiral modes to generate more non-radial kinetic energy than
a single sloshing mode, increasing the size of the average shock radius, and
hence generating better conditions for the formation of large-scale, high-entropy
bubbles. In contrast, convection-dominated explosions show a smaller 
difference in their critical heating rate between 2D and 3D ($<8\%$), 
in agreement with previous studies.
The ability of our numerical implementation to maintain arbitrary symmetries
is quantified with a set of SASI-based tests.
We discuss implications for the diversity of explosion paths in a 
realistic supernova environment.
\end{abstract}

\begin{keywords}
hydrodynamics -- instabilities -- neutrinos 
-- nuclear reactions, nucleosynthesis, abundances -- shock waves -- supernovae: general
\end{keywords}

\section{Introduction}

While the explosion mechanism of massive stars is still
an unsolved problem, important progress has been made 
in the last few decades on the theory side 
(see, e.g., \citealt{foglizzo2015} for a recent review). 
At present, the leading explanation for the majority of core-collapse supernovae is the
\emph{neutrino mechanism}, in which energy deposition by a small fraction of 
the outgoing neutrinos powers the revival of the stalled shock \citep{bethe85}. 
State-of-the-art models have shown that this mechanism fails in spherical symmetry,
except for a small fraction of 
progenitors at the low mass end for core-collapse 
(e.g., \citealt{janka2012a} and referneces therein).

Significant effort has been invested in the development of multi-dimensional hydrodynamic models,
first in axisymmetry (2D; e.g. \citealt{herant92,burrows95,janka96,ott08,suwa2010,bruenn2013}) 
and recently in three spatial dimensions (3D), with 
varying degrees of approximation for other physics such as neutrino transport 
(e.g., \citealt{hanke2013} and references therein).
These studies and others have demonstrated the importance of hydrodynamic instabilities for providing
favorable conditions for explosion.
Observationally, an asymmetric explosion is expected from spectropolarimetry (e.g., \citealt{wang08}),
the distribution of elements in supernova remnants (e.g., \citealt{grefenstette2014}), and
the large proper motions of pulsars (e.g., \citealt{hobbs2005}).

Two types of instability can break the symmetry of a
stalled shock. First, neutrino energy deposition drives buoyant \emph{convection} 
in a layer just inside the shock (e.g., \citealt{bethe90}). This is a local instability 
that generates kinetic energy on relatively small spatial scales (spherical harmonic indices 
$\ell \gtrsim 5$). Second, the shock is unstable to a global
oscillatory instability driven by a cycle of advected and acoustic perturbations
trapped in between the shock and the neutrinosphere: the Standing Accretion 
Shock Instability (SASI; \citealt{BM03,BM06,F07,guilet2012}). This instability favors 
larger spatial scales than convection ($\ell\sim 1-2$). Non-axisymmetric (spiral) modes of the SASI
can also lead to angular momentum redistribution, as found in numerical simulations 
\citep{blondin07a,blondin07b,F10,guilet2014} and in experiments \citep{foglizzo2012}. 

The first generation of 2D supernova models with advanced neutrino transport found strong sloshing activity in 
the shock preceding explosion, and hence the SASI was generally credited for contributing 
to this success (e.g., \citealt{burrows07,marek09}). But because the
inverse cascade in 2D turbulence transfers kinetic energy to large
spatial scales, it was not clear whether these sloshings were
in fact driven by the SASI or just a consequence of convective activity
(e.g., \citealt{FT09b}). Thereafter, the first
batch of 3D hydrodynamic core-collapse models with approximate
neutrino physics did not yield prominent sloshings in exploding cases
\citep{fryer07,nordhaus10a,hanke2012,takiwaki2012,couch2013b,dolence2013},
showing instead all the signs of buoyant convection driving the 
dynamics \citep{burrows2012,murphy2012}.
These early 3D studies however focused on a small set of progenitors.

The full-physics 2D study of \citet{mueller2012} collapsed two progenitors 
($8.1$ and $27M_\sun$) with very different inner density profiles (compactness), 
and found two distinctive explosion paths: one dominated by convection 
and another dominated by the SASI at low and high accretion rates, 
respectively. A followup study of the $27M_\sun$ progenitor in 3D using 
sophisticated neutrino transport found episodic SASI activity, but
no successful explosion \citep{hanke2013}. Other studies using more approximate 
transport did not find a dominant SASI in exploding systems
\citep{ott2013,couch2014,abdikamalov2014}.

The presence of SASI activity leads to specific predictions regarding the
neutrino signal \citep{marek2009b,lund2010,lund2012,tamborra2013,tamborra2014} 
and gravitational wave emission (e.g., \citealt{murphy2009,kotake2011,mueller2013}). 
The differences introduced by dimensionality in \emph{bona fide} SASI-dominated 
explosions have not been systematically explored yet, with the closest work being
the parametric models of \citet{hanke2013} and \citet{iwakami08,iwakami2014}, in which
systems straddle both instability regimes (convection parameter close to the critical value)
at the time when the shock stalls or in the initial condition.
In contrast, multiple 3D studies have addressed convection-dominated
systems, finding either relatively minor differences in the susceptibility to explosion between 
2D and 3D (e.g., \citealt{hanke2012,dolence2013,handy2014}), or more detrimental conditions in 
3D (e.g., \citealt{hanke2013,couch2014,mezzacappa2015,lentz2015}; although recent work
by \citealt{mueller2015b} suggests that 3D can be more favorable than 2D \emph{after} the onset of explosion).
A better understanding of SASI-dominated explosions is important within the 
search for a robust explosion mechanism, because 
the susceptibility to the SASI depends not only on the compactness of the 
progenitor via the accretion rate, 
but also on the speed at which the protoneutron star contracts \citep{scheck08}. 
Therefore, the current generation of models could be underestimating the 
prevalence of the SASI given the treatment of neutrino physics and the uncertainties 
in the stellar models and equation of state (EOS).

Here we investigate the transition to explosion using parameterized
hydrodynamic models that, while approximate, are free from
uncertainties about the progenitor and the EOS, allowing a
clean test of the hydrodynamic effects in 2D and 3D. Model parameters can be controlled so 
that at explosion, they lie well within the regime of dominance of either instability. 
Our approach builds on the study of \citet[ hereafter Paper I]{FMFJ14}, who 
characterized the properties of SASI- and convection-dominated explosions in 
2D. The main conclusion of that study is that both types of 
explosion succeed by generating large-scale, high-entropy bubbles in different ways:
either directly via large-scale shock expansions (SASI) or 
through the dynamics of convective bubble growth.
Our focus here is to probe the differences introduced by adding the 3rd spatial dimension
to SASI-dominated systems, and to
compare the outcome to convection-dominated explosions.

The structure of the paper is the following. Section~\ref{s:methods} describes the physical
model and numerical method used, and the list of models evolved. Section~\ref{s:results}
presents our results, with an overview of the transition to explosion,
the role of dimensionality and resolution, and the properties
of non-exploding SASI-dominated systems. Section~\ref{s:summary} summarizes our
results and discusses the implications for our understanding of
core-collapse supernovae. The Appendix describes the extension of the
code to 3D and a set of SASI-based tests to quantify its accuracy.

\section{Methods}
\label{s:methods}
\subsection{Physical Model and Initial Condition}
\label{s:initial_condition}

In order to study the hydrodynamic instabilities that arise during
the stalled shock phase of core-collapse supernovae,
we employ a non-rotating spherical accretion flow
with a standing shock as the baseline state. This solution
assumes constant mass accretion rate, constant neutron star
radius, and employs a gamma-law equation of state with parameterized
charged-current neutrino source terms, including `light bulb' irradiation.
This type of model was first used by \citet{BM03} without heating to isolate the SASI,
and has been subsequently employed in a number of parametric
studies of the CCSNe mechanism \citep{foglizzo06,F07,FT09a,FT09b,guilet2010,guilet2012}. 

This parametric approach has two advantages:
(1) the results are independent of the uncertainties
in stellar models, the dense matter EOS, or an incomplete treatment
of neutrino transport; and
(2) the linear stability properties of the system 
are well understood \citep{F07,FT09a},
enabling the construction of model sequences in well-defined regions
of parameter space. 
The resulting flow in the gain
region compares favorably with models that include more physics (Paper I).

The initial condition consists of a steady-state spherical 
accretion flow incident on a neutron star of mass $M_{\rm ns}$ 
and radius $r_*$, with a standing shock at a radius $r_s$. The flow 
outside of the shock is adiabatic and supersonic, with Mach number 
$\mathcal{M}_0$ at a fiducial radius $r_0$.
Inside the shock the flow is subsonic, decelerating onto the neutron
star surface through the emission of neutrinos and geometric convergence. 
The specific rate of neutrino heating and cooling is parameterized as
\begin{equation}
\label{eq:neutrino_source}
Q_\nu = \left[\frac{B}{r^2} - Ap^{3/2} \right]\, e^{-(s/s_{\rm min})^2}\,\Theta(\mathcal{M}_{\rm cut}-\mathcal{M}),
\end{equation}
where $p$ is the pressure and $s$ is the entropy. The minimum entropy $s_{\rm min}$ prevents runaway
cooling, while the cutoff Mach number $\mathcal{M}_{\rm cut}$ suppresses source terms in the upstream flow.

To generate a model sequence, we first fix the cooling normalization in the
absence of heating ($B=0$).  Given the ratio of shock to stellar radius
$r_s/r_*$, with $r_s = r_0$, the adiabatic index $\gamma$, and the constant rate of
nuclear dissociation at the shock $\varepsilon$, the cooling normalization $A$ is found by
requiring that the radial velocity vanishes at $r=r_*$. The sequence is then
formed by varying the heating constant $B$, which is related to the lightbulb neutrino 
luminosity by
\begin{equation}
\label{eq:B_dimensional}
B \simeq 0.007\, L_{\nu_e,52}\, T_{\nu,4}^2\,
\left(\frac{r_{\rm 0}}{100\textrm{ km}}\right)^{1/2}\left(\frac{1.3M_\odot}{M_{\rm ns}} \right)^{3/2},
\end{equation}
where $L_{\nu_e,52}=L_{\nu_e}/(10^{52}\textrm{erg s}^{-1})$ is the electron neutrino
luminosity and $T_{\nu,4} = T_{\nu}/(4\textrm{ MeV})$ is the neutrinospheric temperature.
Increasing $B$ increases the initial shock radius $r_s$ from its value without heating, $r_0$. The
minimum entropy $s_{\rm min}$ is obtained separately for each model, by computing the entropy
at $r=r_*$ without using the exponential cutoff in equation~(\ref{eq:neutrino_source}). 
The numerical value is obtained using the ideal gas entropy normalized to its 
post-shock value (e.g., \citealt{F07})
\begin{equation}
\label{eq:entropy_definition}
s = \frac{1}{\gamma-1}\ln\left[\frac{p}{p_2}\left(\frac{\rho_2}{\rho}\right)^\gamma\right],
\end{equation}
where $\rho_2$ and $p_2$ are the initial post-shock density and pressure.
The initial condition for the time-dependent simulation is then obtained by 
re-computing the solution including the entropy cutoff $s_{\rm min}$ 
in equation~(\ref{eq:neutrino_source}).

To connect with previous work, we employ an adiabatic index $\gamma=4/3$, a
ratio of shock to stellar radius without heating $r_0/r_* = 0.4$, and an
upstream Mach number $\mathcal{M}_0 = 5$. Examples of the type of initial condition
obtained with this model can be found in \citet{FT09b}. Throughout the
paper, we make use of a general unit system that reflects the dimensionless
character of the problem. These units and their characteristic 
values in the core-collapse supernova problem are described in Table~\ref{t:units}.

\begin{table}
\centering
\begin{minipage}{8cm}
\caption{Frequently-used quantities and reference values applicable to 
core-collapse supernovae. The use of a gamma-law equation of state 
and an arbitrary normalization for the cooling function allow a
dimensionless formulation of the problem. \label{t:units}} 
\begin{tabular}{lp{2.5cm}l}
\hline
{Symbol} &
{Quantity} &
{CCSN Reference value}\\
\hline
\noalign{\smallskip}
$r_0$        & Initial shock radius with $B=0$ & $100\,r_{\rm 100}$~km\\
$B$          & Heating amplitude            & Equation~(\ref{eq:B_dimensional})\\
\noalign{\smallskip}
$M_{\rm ns}$ & NS mass                      & $1.3\,M_{1.3}$~$M_\sun$\\
\noalign{\smallskip}
$\dot M$     & Accretion rate               & $0.2\dot{M}_{0.2}$~$M_\sun$~s$^{-1}$\\
\noalign{\smallskip}
$v_{\rm ff}$ & $(2GM_{\rm ns}/r_0)^{1/2}$   & $10^{9.77}\,M_{\rm 1.3}^{1/2}r_{\rm 100}^{-1/2}$~cm~s$^{-1}$\\
\noalign{\smallskip}
$t_0$        & $r_0/v_{\rm ff}$             & $1.7\,r_{\rm 100}^{3/2}M_{1.3}^{-1/2}$~ms\\
\noalign{\smallskip}
$E_0$        & $\dot{M} r_0 v_{\rm ff}$     & $10^{49.37}\dot{M}_{0.2}r^{1/2}_{\rm 100}M^{1/2}_{1.3}$~erg\\
\hline
\end{tabular}
\end{minipage}
\end{table}

\subsection{Numerical Setup}
\label{s:setup}

We use {\textsc FLASH3.2} \citep{fryxell00,dubey2009} to solve
the Euler equations in spherical polar coordinates $(r,\theta,\phi)$,
subject to the gravity of a point mass and neutrino energy source terms,
\begin{eqnarray}
\frac{\partial \rho}{\partial t} + \nabla\cdot(\rho\mathbf{v}) & = & 0\\
\frac{\partial \mathbf{v}}{\partial t} + (\mathbf{v}\cdot \nabla)\mathbf{v}
                                     &  = & -\frac{1}{\rho}\nabla p - \frac{GM_{\rm ns}}{r^2}\hat r\\
\frac{\totd e_{\rm int}}{\totd t} -\frac{p}{\rho^2}\frac{\totd \rho}{\totd t} & = & Q_\nu.
\end{eqnarray}
where $\rho$, $\mathbf{v}$, $e_{\rm int}$ are the density, velocity,
and internal energy, and 
$\totd/\totd t = \partial/\partial t + \mathbf{v}\cdot \nabla$. The system
of equations is closed with an ideal gas equation of state, and the point
mass $M_{\rm ns}$ remains constant in time.

The public version of FLASH has been modified to enable the split
Piecewise Parabolic Method (PPM) solver to operate in 3D spherical coordinates. A detailed
description of the changes to the code and verification tests 
of the implementation are presented in Appendix~\ref{s:code_tests}. 
The code has previously been modified to enable use of a non-uniform 
grid, as described in \citet{F12}.

The computational domain covers the full range of polar angles, $\theta\in[0,\pi]$
and $\phi \in [0,2\pi]$. In the radial direction, the domain extends
from $r=r_*=0.4r_0$ to $r=7r_0$. The radial grid is logarithmically
spaced with $640$ cells, resulting in a constant fractional 
spacing $\Delta r / r = 0.45\%$ or $0.26^\circ$. The meridional grid has
constant spacing in $\cos\theta$, and the azimuthal grid is uniform.
This angular spacing yields cells that subtend a constant solid
angle at fixed radius.

We adopt two resolutions for the angular grid in order to test the 
sensitivity of the results to this variable. The baseline resolution
employs $56$ cells in the $\theta$ direction and $192$ cells in $\phi$,
for an effective resolution $\Delta \theta \simeq \Delta \phi \simeq 2^\circ$
at the equator. In addition, we evolve two models at twice the resolution
in both $\theta$ and $\phi$ ($112$ and $384$ grid points, respectively). 
Our radial resolution is among the highest in published 3D hydrodynamic supernova studies,
while our baseline angular resolution is comparable to that of 
\citet{hanke2013}, \citet{mezzacappa2015}, and \citet{melson2015}. Our 
highest angular resolution ($1^\circ$ at the equator) is only a factor of 2 lower
than the highest resolution of \citet{burrows2012}, \citet{couch2014}, and
\citet{abdikamalov2014}.

The boundary condition in azimuth is periodic. In the radial direction,
we use a reflecting inner boundary at $r=r_*$,
and set the ghost cells to the steady-state solution at the outer boundary. 
In the $\theta$ direction, we fill the ghost cells with information from active
cells located across the axis (Appendix~\ref{s:code_tests}); for comparison, we also 
run a model with a reflecting boundary condition in $\theta$. 

\subsection{Models Evolved}
\label{s:models}

We evolve two sequences of models, each with increasing heating rate 
$B$ (eq.~[\ref{eq:B_dimensional}]), as shown
in Table~\ref{t:models}. The two sequences differ in the magnitude of 
the nuclear dissociation $\varepsilon$ at the shock, which changes the 
strength of the shock jump and thus the magnitude of the postshock velocity 
(e.g., \citealt{thompson00}). 
This difference places each sequence well above and below the bifurcation
in parameter space that determines whether the SASI or convection dominate the dynamics.

The relative importance of each instability is set by the relation between
the advection rate and convective growth rate in the initial
condition. This can be quantified with
the `convection' parameter \citep{foglizzo06}
\begin{equation}
\label{eq:chi_definition}
\chi = \int_{r_g}^{r_s} \frac{|\omega_{\rm BV}| \totd r}{|v_r|},
\end{equation}
where $\omega_{\rm BV}$ is the Brunt-V\"ais\"al\"a frequency, and
the integral runs from the gain radius $r_g$ to the shock
radius $r_s$. The transition from SASI- to convection-dominance
occurs around $\chi \simeq 3$.
Our \emph{SASI-dominated} sequence (models names starting with `s') 
has $\varepsilon=0$, yielding values of $\chi \ll 3$ in all initial models, 
while the \emph{convection-dominated} sequence (model names starting with `c') 
has $\varepsilon = 0.3GM_{\rm ns}/r_0$, yielding  $\chi \gg 3$ for 
all values of $B$.

For both sequences, we evolve models in 1D, 2D, and 3D around the 
critical value of $B$ for which an explosion is obtained. For 1D and 2D,
these critical values were obtained in Paper I. Here we 
recompute the same models to maintain consistency with 3D in the 
numerical implementation and grid resolution\footnote{The models of Paper I did not
employ a hybrid Riemann solver at the shock, whereas our 3D implementation
requires such a hybrid approach to minimize numerical noise (Appendix~\ref{s:code_tests}).}. The 
critical heating rates for 1D and 2D models are consistent with the previous value.

Two types of initial perturbation are applied. In most cases, random
velocity perturbations in all coordinate directions are applied over the
entire computational domain at $t=0$, with magnitude $0.1\%$ of the radial velocity.
In a few models, an overdense spherical shell with an $\ell=1$ angular
dependence is added to the flow (as in, e.g., \citealt{BM06} or \citealt{FT09a}). 
The dipole axis of this perturbation is not aligned with any cartesian coordinate direction, 
pointing instead towards $\theta = \phi = 45^\circ$. This perturbation excites a clean
sloshing SASI mode, thus imposing a deterministic initial condition for probing the effects
of dimensionality and resolution. Appendix~\ref{s:code_tests} shows how this type of perturbation
is used as a test of the ability of the code to maintain different symmetries.

A model is considered to have exploded when the shock radius reaches or approaches
the outer radial computational boundary. For the 1D case with finite nuclear dissociation,
we assign an explosion time at the moment when the periodic oscillation cycle is broken.
Explosion times are listed in Table~\ref{t:models}. Most models are evolved up to a time 
$t=500t_0$ or until they explode, whichever comes first\footnote{The high-resolution
model S08L1-hr is interrupted earlier than $500t_0$ when it appears that it will
not explode, to save computing time.}.

\begin{table}
\centering
\begin{minipage}{8cm}
\caption{Model parameters and explosion time.\label{t:models}
Columns show from left to right: model
name, nuclear dissociation energy at the shock, heating normalization 
(eqns.~[\ref{eq:neutrino_source}]-[\ref{eq:B_dimensional}]), initial $\chi$ parameter (eq.~[\ref{eq:chi_definition}]),
type of perturbation ($\delta\mathbf{v}$: random velocity, $\ell=1$: overdense shell), and explosion time.
A model is considered to have exploded when the shock hits or approaches the outer 
simulation boundary. For 1D cases with $\varepsilon\ne 0$, 
the model `explodes' when the shock oscillation is broken by transient
expansion (e.g. model c1d-08).}
\begin{tabular}{lccccc}
\hline
{Model} &
{$\varepsilon$} &
{$B$}   &
{$\chi$} &
{Pert.} &
{$t_{\rm exp}$} \\
{} & {$(GM_{\rm ns}/r_0)$} & {$(10^{-3})$} & {} & {} & {$(t_0)$}\\
\hline
\noalign{\smallskip}
\noalign{1D:}
\noalign{\smallskip}
s1d-09 &  0  &  9 & 1.3 & ... & ... \\
s1d-10 &     & 10 & 1.6 & ... & ... \\
s1d-11 &     & 11 & 1.8 & ... & 314 \\
\noalign{\smallskip}
c1d-07 & 0.3 & 7  & 9.1 & ... & ... \\
c1d-08 &     & 8  & 11  & ... & 350\\
\noalign{\smallskip}
\noalign{\smallskip}
\noalign{2D:}
\noalign{\smallskip}
s2d-00   & 0   &  0  &  0   & $\delta \mathbf{v}$ & ... \\
s2d-06   &     &  6  & 0.6  &                     & ... \\
s2d-08   &     &  8  & 1.1  & $\ell=1$            & ... \\
s2d-09   &     &  9  & 1.3  & $\delta \mathbf{v}$ & ... \\
s2d-09hr &     &     &      &                     & ... \\
s2d-10   &     & 10  & 1.6  &                     & 238 \\
s2d-10hr &     &     &      &                     & 256 \\
\noalign{\smallskip}
c2d-06    & 0.3 &  6  & 7.1  &                     & ... \\
c2d-06hr  &     &     &      &                     & ... \\
c2d-065   &     & 6.5 & 8.0  &                     & ... \\
c2d-065hr &     &     &      &                     & ... \\
c2d-07    &     &  7  & 9.1  &                     & 423 \\
c2d-07hr  &     &     &      &                     & 339 \\
\noalign{\smallskip}
\noalign{\smallskip}
\noalign{3D:}
\noalign{\smallskip}
S00dv  & 0 &  0 &  0  & $\delta \mathbf{v}$  & ... \\ 
S06dv  &   &  6 & 0.6 &                      & ... \\ 
S07dv  &   &  7 & 0.8 &                      & ... \\ 
S08dv  &   &  8 & 1.1 &                      & 306 \\ 
S09dv  &   &  9 & 1.3 &                      & 204 \\ 
S10dv  &   & 10 & 1.6 &                      & 219 \\ 
\noalign{\smallskip}
S08L1     & 0 & 8 & 1.1 & $\ell=1$           & 172 \\
S08L1-ref &   &   &      &                    & 270 \\
S08L1-hr  &   &   &      &                    & ... \\  
S09dv-hr  &   & 9 & 1.3 & $\delta\mathbf{v}$ & 187 \\
\noalign{\smallskip}
C06dv     & 0.3  & 6 & 7.1 & $\delta\mathbf{v}$ & ...\\
C065dv    &      & 7 & 8.0 &                    & ...\\
C07dv     &      & 7 & 9.1 &                    & 336\\
C07dv-hr  &      &   &      &                    & 313\\ 
\hline
\end{tabular}
\end{minipage}
\end{table}

\begin{figure*}
\includegraphics*[width=\textwidth]{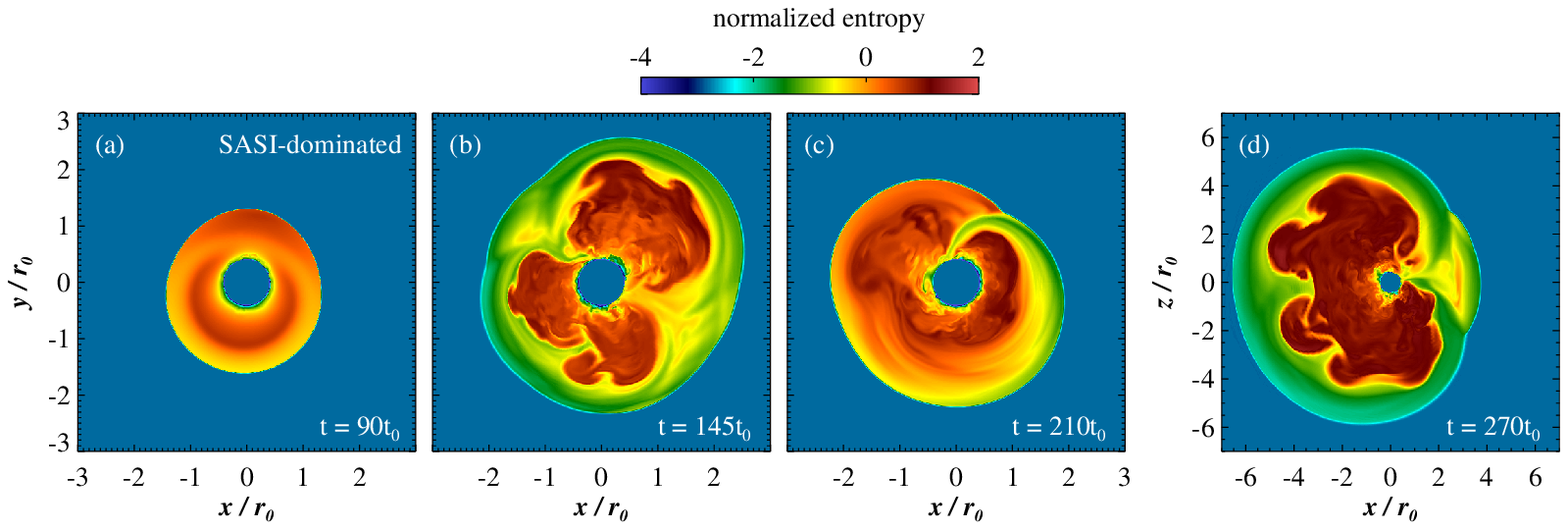}
\includegraphics*[width=\textwidth]{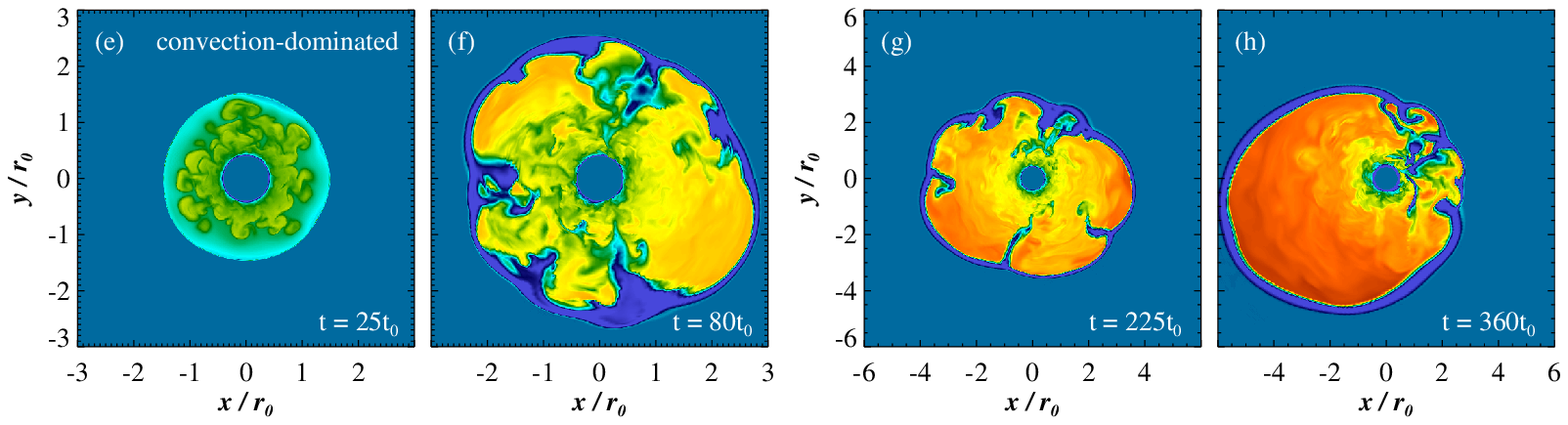}
\caption{Snapshots in the evolution of the marginally exploding SASI-dominated model (S08dv, top) and
convection-dominated model (C07dv, bottom), both with random initial velocity perturbations. Each panel shows a slice
of the entropy (eq.~[\ref{eq:entropy_definition}]) on the $xy$ plane; to ease comparison,
the entropy in model C07dv is normalized to the initial postshock value of model S08dv.
The SASI-dominated model initially develops a sloshing mode (panel a) which saturates (panel b). The subsequent
development of a spiral mode (panel c) generates a large-scale, high-entropy bubble which leads
to the final runaway (panel d). The convection-dominated model initially develops small convective
plumes (panel e) which gradually grow into larger structures (panels f-g), until a single dominant
bubble remains (panel h). Note the different spatial scale in panels (d), (g), and (h).}
\label{f:entropy_slices_explosion}
\end{figure*}

\begin{figure*}
\begin{overpic}[width=\columnwidth,clip=true]{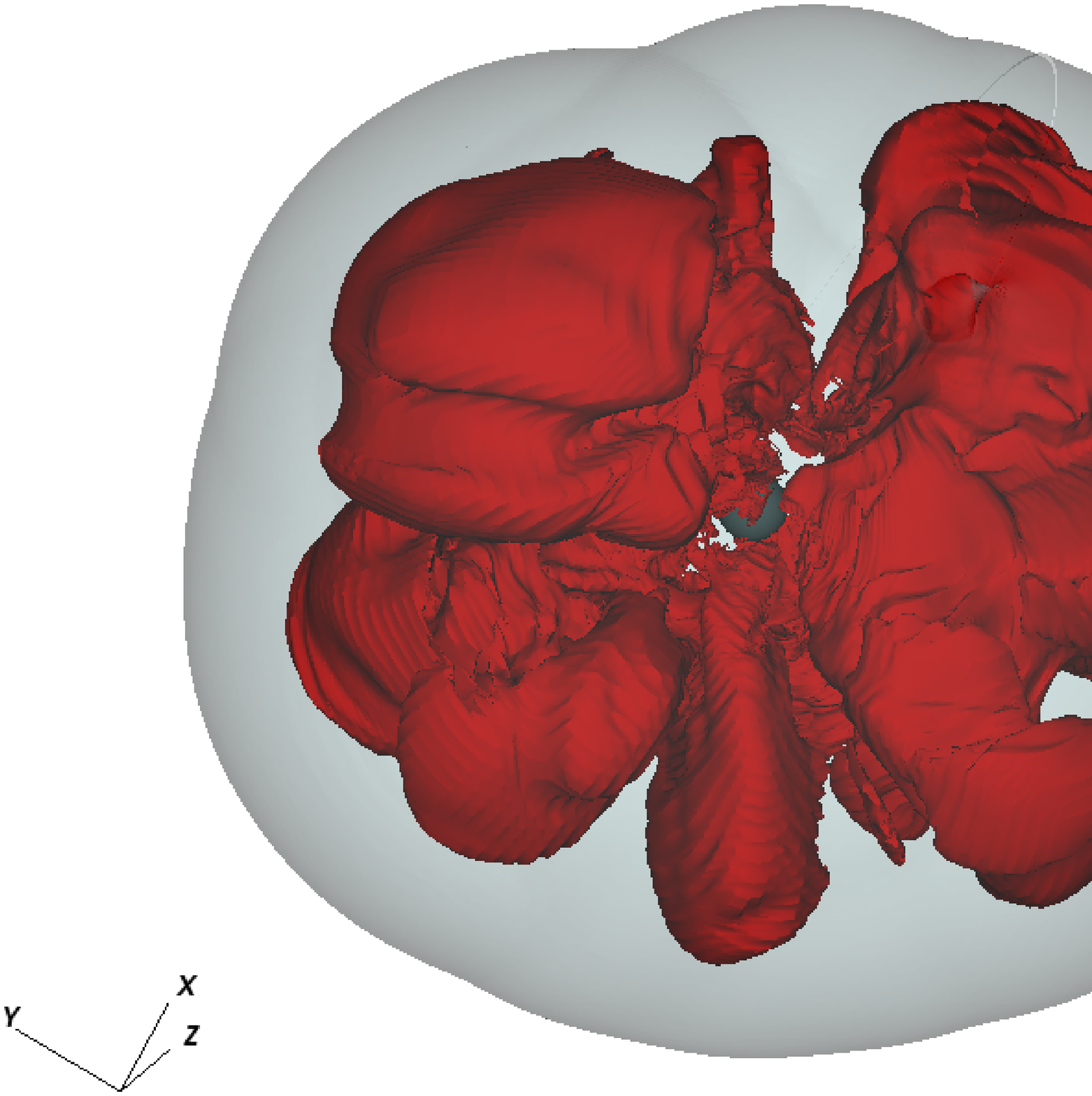}
\put(35,90){{\large S08dv (SASI-dominated)}}
\end{overpic}
\begin{overpic}[width=\columnwidth,clip=true]{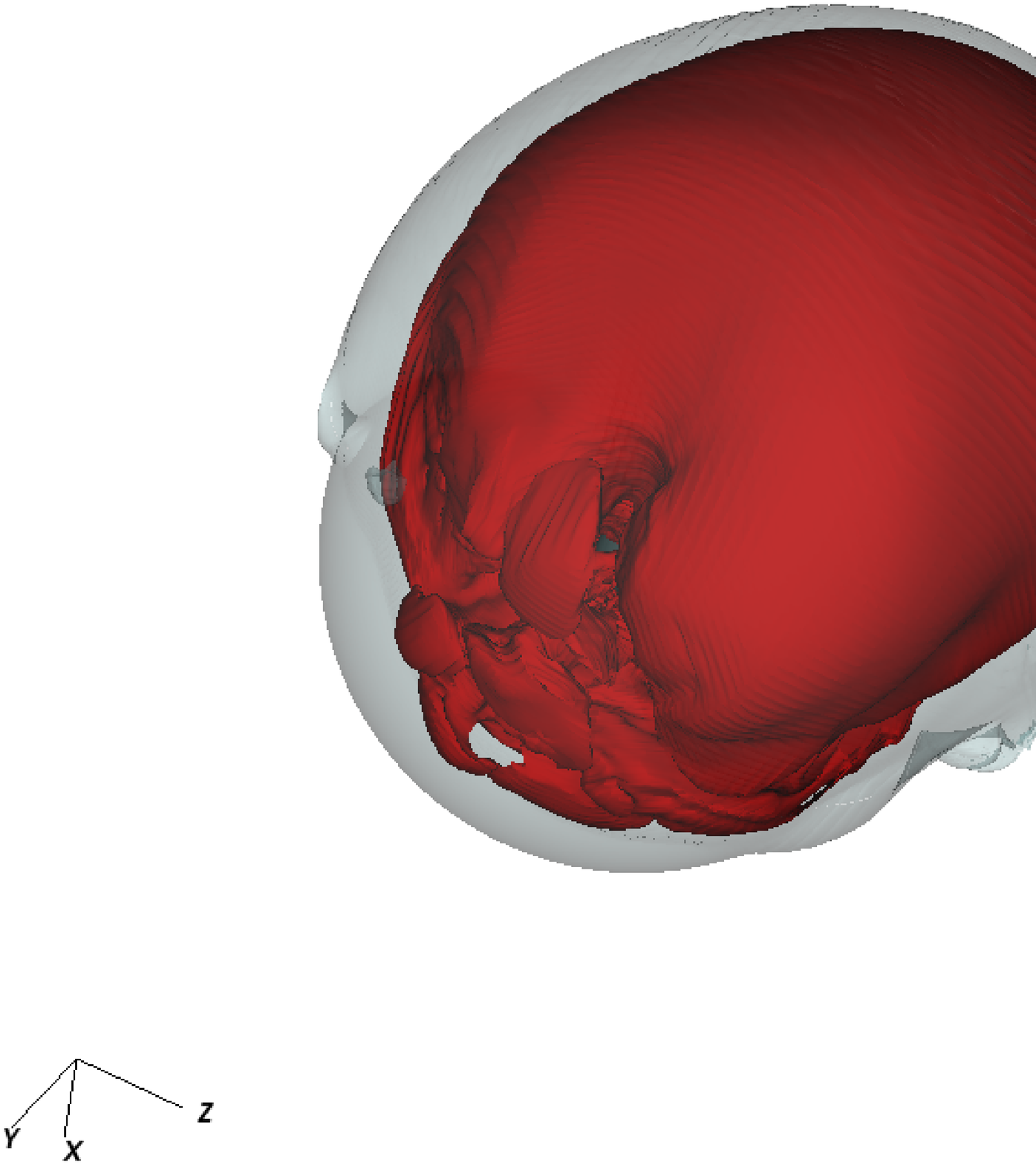}
\put(25,90){{\large C07dv (Convection-dominated)}}
\end{overpic}
\caption{Explosion geometry near the end of the simulation for the marginally exploding 
SASI-dominated model (S08dv, left) and convection-dominated model (C07dv, right), corresponding
to panels (d) and (h) of Figure~\ref{f:entropy_slices_explosion}. 
A representative entropy isosurface is shown in red, the shock is the outer 
light grey surface, and the neutron star is the dark grey sphere (the scale is different in the two panels).}
\label{f:entropy_shock_3d_sasi}
\end{figure*}

\section{Results}
\label{s:results}

\subsection{Transition to Explosion: Overview}
\label{s:transition}

To illustrate the general behavior of SASI-dominated systems, we describe
the evolution of the marginally-exploding 3D model S08dv (Table~\ref{t:models}).
Initially, a dominant sloshing SASI mode grows out of random initial 
velocity perturbations, as shown in Figure~\ref{f:entropy_slices_explosion}a.
As this mode achieves saturation, sloshing modes in 
orthogonal directions also grow and
become non-linear, slightly out-of-phase, but the resulting spiral mode\footnote{ 
\citet{F10} showed that spiral modes can be described as a superposition of linearly
independent sloshing modes (quantified by the real spherical harmonic coefficients 
of the shock) which are out
of phase relative to each other. This phase difference does not need to be $\pi/2$.} 
survives only for a fraction of an oscillation period and the shock expansion dies down. 

After this lull in shock expansion around $t\sim 200t_0$, a coherent spiral mode arises, growing
quickly in magnitude for three full oscillations. By this time, a large high-entropy
bubble has developed, and the triple-point intersects this bubble, interrupting the
oscillation. Runaway does not follow immediately, however, with the bubble lingering
for a period $\sim 30t_0$, which allows turbulence to partially break the bubble down into
smaller parts.
This accounts for the multiple plumes seen in Figure~\ref{f:entropy_slices_explosion}d,
which become more apparent when visualizing the final explosion geometry
in Figure~\ref{f:entropy_shock_3d_sasi}. If a realistic EOS that includes 
the energy release from alpha particle
recombination had been used, the explosion pattern would have likely frozen at a smaller 
radius, and the geometry would consist of a single dominant bubble.

The development of SASI activity is an optimal way to rapidly seed large-scale
high-entropy bubbles. Whenever the shock expands relative to its equilibrium
position, the Mach number in the frame of the shock increases, and the
entropy jump correspondingly also increases (e.g., Paper I). The formation
of such large-scale bubbles in 3D was also seen by \citet{hanke2013} when the
SASI was active. 

The interplay between bubbles and SASI oscillations
is also present in 2D (Paper I). In that case, it is found
that large bubbles cut off accretion to the cooling region, which 
in turn breaks the SASI oscillation cycle. If bubbles are able to survive
for a time longer than approximately a few SASI oscillations, 
a runaway expansion of the shock
will ensue. In 3D, the same dynamics appears to take place,
but now bubbles can have larger sizes 
given the lack of an axisymmetry constraint. 
As in in 2D, failure to achieve runaway expansion
in 3D appears to be related to  
the disruption of bubbles due to either accretion plumes and/or 
turbulence. We elaborate on these processes in \S\ref{s:dim_res} 
and \S\ref{s:non_exploding}.

The dynamics of convection-dominated explosions is very different.
At early times, numerous small convective plumes grow out of the random
initial perturbations and 
fill the gain region, as shown in Figure~\ref{f:entropy_slices_explosion}.
The added convective stresses cause the shock to gradually expand,
with convective plumes 
growing in size. This 
process continues until there is a single, dominant convective
bubble that approaches the edge of the computational domain. 

\begin{figure*}
\includegraphics*[width=\columnwidth]{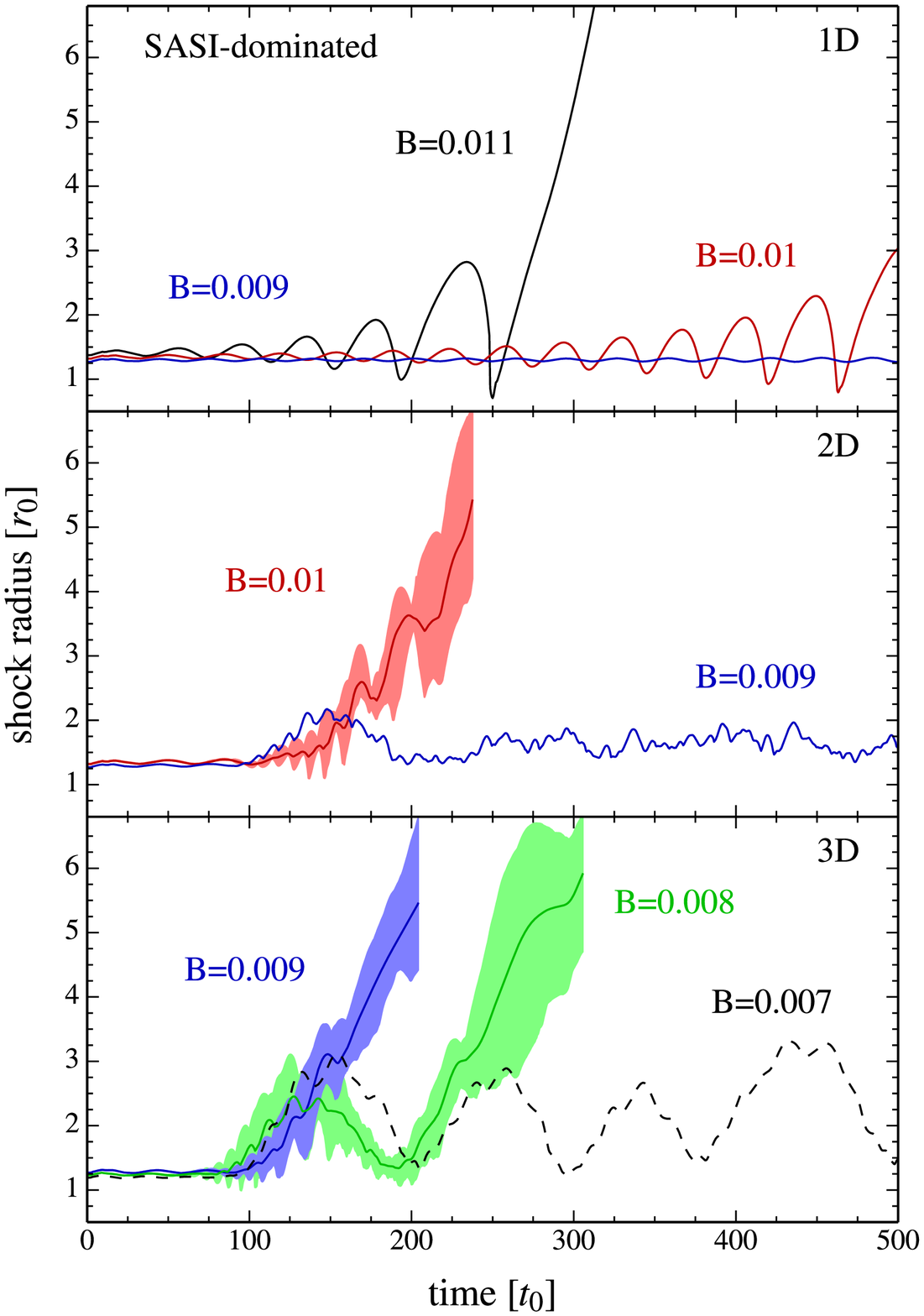}
\includegraphics*[width=\columnwidth]{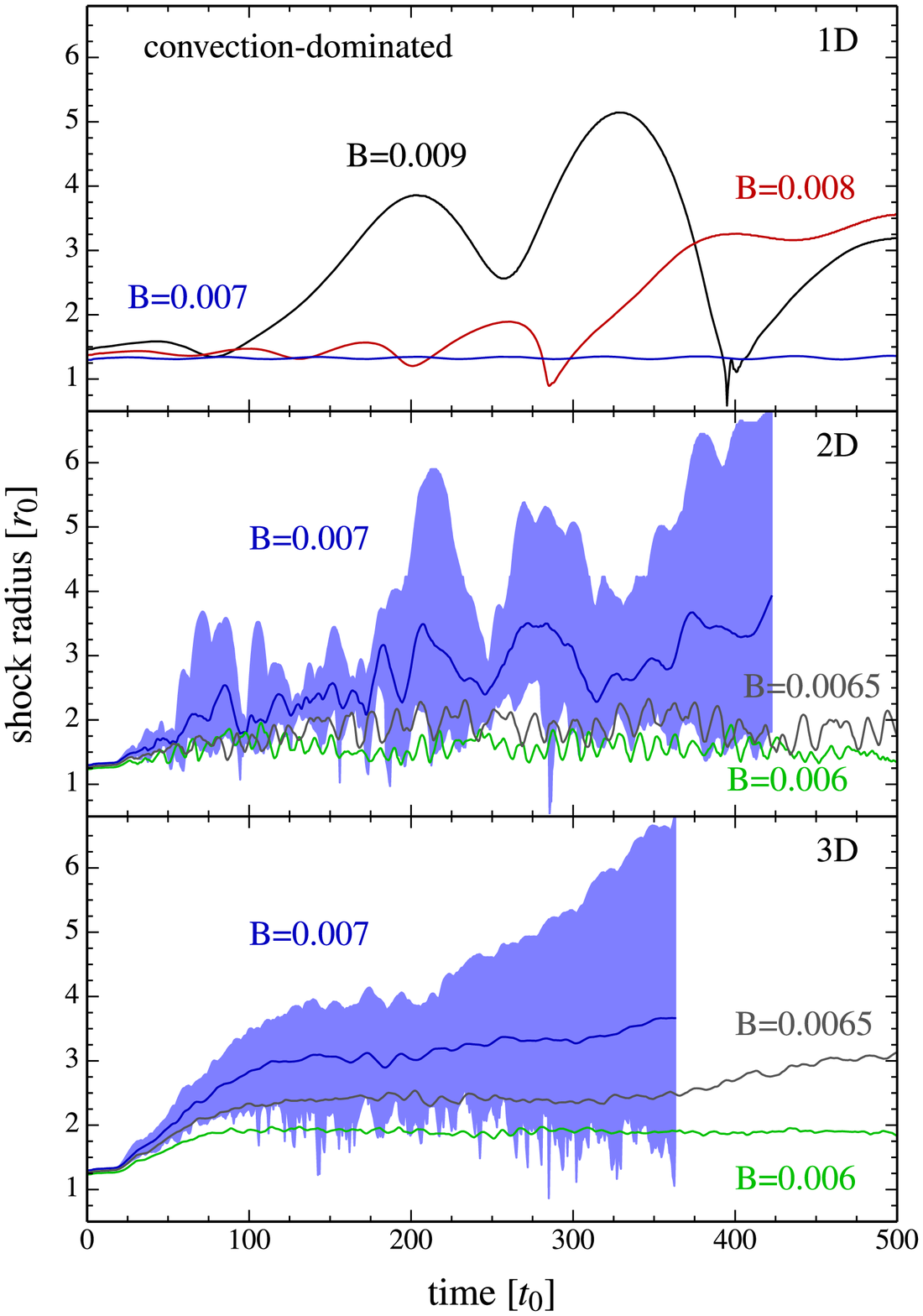}
\caption{Evolution of the shock radius for SASI-dominated models (left) 
and convection-dominated models (right) with random initial velocity perturbations. 
Top, middle, and bottom panels show models around the critical transition to 
explosion in 1D, 2D, and 3D, respectively. For 2D and 3D models, solid lines denote the 
angle-averaged shock radius, and the shaded area marks the region between the minimum 
and maximum shock radii. Note that 3D SASI-dominated models can explode at a lower neutrino
luminosity than in 2D, and that non-exploding SASI-dominated models
undergo much larger shock excursions in 3D than in 2D (\S\ref{s:non_exploding}).}
\label{f:shock_critical}
\end{figure*}

The increase in the shock radius due to 
convective stresses has been studied extensively 
(e.g., \citealt{murphy2012,couch2015a}). The formation of a dominant 
structure as the limiting case of bubble growth and/or consolidation
is an intrinsic property of the non-linear Rayleigh-Taylor instability 
\citep{sharp1984}. A characteristic bubble size that subtends a solid angle
$\sim 1$~radian was predicted theoretically by \citet{thompson00} as
a result of a balance of buoyancy and ram pressure on the bubble, and
has been further studied by \citet{dolence2013} and \citet{couch2013b}.

The slow growth of this bubble is in part an artifact of the simplified
nuclear dissociation prescription, which causes an excess energy
loss relative to the gravitational potential energy as the shock moves out 
(e.g., \citealt{FT09b}). 
With a more complete EOS,
runaway would have ensued
at a smaller radius and hence the convective pattern would have
frozen before reaching this stage (\ref{f:entropy_shock_3d_sasi}). 
Nevertheless, it is interesting to note 
that this geometry, with a one-sided
bubble, has also been observed in a simulations 
that include the full equation of state \citep{dolence2013,lentz2015}.

\subsection{Dependence on Dimensionality and Resolution}
\label{s:dim_res}

The most interesting result of this investigation is that 
SASI-dominated explosions can take place at lower neutrino luminosities
with increasing dimensionality. Figure~\ref{f:shock_critical} illustrates the 
magnitude of the
effect, showing the evolution of the shock radius with time for 
systems around the critical heating rate. Models in 3D
can explode at neutrino luminosities $\sim 20\%$ lower than in 2D.

In contrast, convection-dominated systems show a much smaller
difference in the critical luminosity between 2D and 3D within
the spacing in $B$ studied here ($\sim 8\%$). The underlying
difference will 
manifest itself at smaller intervals of 
$B$, but at that point the result will also depend on other
factors such as the numerical resolution or the detailed form of 
pre-collapse perturbations (e.g., \citealt{couch2015a,mueller2015}). 

Most of the work that has analyzed the dependence of 2D-3D differences
in convection-dominated systems has concluded that higher
resolution is detrimental for 3D 
(e.g., \citealt{hanke2012,couch2013b,takiwaki2014}). 
This trend has been attributed to the direction in which
kinetic energy flows in the turbulent cascades in 2D and 
3D \citep{hanke2012}. 
In our models, simply doubling the angular resolution does not prove to
be very informative. The marginal 2D model at high resolution (c2d-07hr) 
explodes significantly earlier, as expected, but the explosion time of the 
marginal 3D model (C07dv-hr) is not very different\footnote{Small variations
in the explosion time are likely affected by stochastic fluctuations, and
should not be taken as a primary indicator for susceptibility to explosion.}.
The non-exploding 3D model with the highest heating rate (C065dv) 
appears to be starting an explosion on a long timescale, but due to the lack 
of a clear runaway we consider it a failure.
Given the significant body of knowledge on this type of 
explosions, we will not concern ourselves further with convection-dominated 
systems.

\begin{figure}
\includegraphics*[width=\columnwidth]{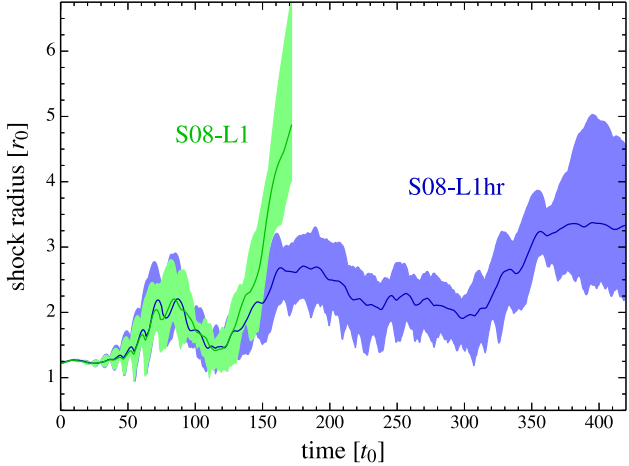}
\caption{Evolution of the shock radius for two SASI-dominated models that
differ only in their angular resolution. Solid lines show the angle-averaged shock
radius, with the shaded area marking the space between minimum and maximum shock radius.
In both cases, an $\ell=1$ sloshing
SASI mode is initially excited at an angle to the coordinate axes ($\theta=\phi=45^\circ$).
Model S08L1 is just above the critical heating rate for explosion. Doubling the
resolution in both $\theta$ and $\phi$ (model S08L1-hr) leads to a failure 
within the simulated time.}
\label{f:shock_critical_resolution}
\end{figure}

The \emph{existence} of a decrease in the critical luminosity 
with increasing dimensionality for SASI-dominated systems is 
not very sensitive 
to resolution, as inferred from the fact that the model
with $B=0.009$ explodes with
standard (S09dv) and high angular resolution (S09dv-hr),
at times within $\sim 10\%$ of each other, and with a heating rate that is 
already below the critical value in 2D. Nevertheless, the \emph{magnitude} 
of this improvement appears to decrease with resolution, as found from 
the marginally-exploding model S08L1. When doubling the angular 
resolution in $\theta$ and $\phi$, the model does not explode, even though 
very large shock excursions still occur (Figure~\ref{f:shock_critical_resolution}). 
If the full EOS was employed, this model would likely have exploded due 
to the additional energy from alpha particle recombination at large radius, 
but within the approximations made in this study we consider it a failure.
SASI-dominated systems with heating rates slightly below the critical 
value display much larger shock excursions in 3D than in 2D, a result
that could potentially lead to further improvement in explosion conditions once
the full EOS is included (c.f. \S\ref{s:non_exploding}). Finally, the
decrease in the critical luminosity is independent on the boundary
condition employed at the polar axis: model S08dv-ref, which employs
a reflecting boundary condition, explodes at a time similar to the
fiducial model S08dv.

We now explore quantitatively the effects of dimensionality and 
resolution in SASI-dominated models. As a baseline
of comparison, we take the marginally-exploding model in 
3D for which an initial $\ell=1$ sloshing SASI mode
is excited via an overdense shell (S08L1). This deterministic
initial perturbation allows a straightforward comparison with 
an identical model in 2D (s2d-08L1), and another in
3D with twice the resolution in both $\theta$ and $\phi$ (S08L1-hr).

A number of time-dependent diagnostics are employed in the
analysis. Aside from the angle-averaged shock radius, we
compute the real $\ell=1$ spherical harmonic coefficients in three
orthogonal (Cartesian) directions:
\begin{equation}
\label{eq:a_i}
a_i(t) = \int \totd \Omega\, Y_i(\theta,\phi)\, r_s(\theta,\phi,t),
\end{equation}
where $r_s$ is the instantaneous shock surface, obtained as a
pressure-gradient-weighted average for smoothness, and the real
spherical harmonics basis functions are
\begin{equation}
\label{eq:real_spherical_harmonics}
Y_{\{x,y,z\}} = \sqrt{\frac{3}{4\pi}}\left\{\sin\theta\cos\phi,\,\sin\theta\sin\phi,\,\cos\theta\right\}.
\end{equation}
The resulting coefficients can be combined linearly in order to
measure sloshing modes along arbitrary axes\footnote{In the 3D
models studied here, a sloshing mode is initially excited along
$\theta=\phi=45^\circ$. Re-defining this direction as the new $z$ axis,
the real spherical harmonic coefficients transform according to: 
\begin{eqnarray}
a_z & \to & 0.5(a_x+a_y)+a_z/\sqrt{2},\nonumber\\
a_x & \to & 0.5(a_x+a_y)-a_z/\sqrt{2},\nonumber\\
a_y & \to & (a_y-a_x)/\sqrt{2}.\nonumber
\end{eqnarray}}. 
Sloshing SASI modes manifest as sinusoidal oscillations in these coefficients.
Multiple coefficients oscillating in phase indicate sloshing
modes along directions not aligned with the coordinate axes.
Out-of-phase oscillations between different coefficients are
the signature of spiral SASI modes \citep{F10}. 
The phase difference does not have to be $\pi/2$ for the
composite mode to display the characteristic properties of a spiral mode (i.e., angular
momentum redistribution). We have ignored modes with $\ell\geq 2$
given that $\ell=1$ is expected to be the most unstable oscillatory
SASI mode at any heating rate (e.g., \citealt{FT09a}; Paper I) and to
keep the analysis concise.

We diagnose the presence of bubbles with enhanced entropy
in the post-shock domain by computing the instantaneous
fraction of the postshock volume $V$ occupied by fluid with
entropy higher than a suitably chosen value $s_0$ (Paper I):
\begin{equation}
\label{eq:fv_definition}
f_V = \frac{1}{V}\int_{s_0}^{s_{\rm max}} \frac{\totd V}{\totd s}\totd s.
\end{equation}
The creation, growth, and destruction of bubbles can be 
characterized quantitatively by following the evolution of
$f_V$ evaluated at multiple values of $s_0$.

In addition, we compute the kinetic energies in the fluctuating
components of the flow
\begin{eqnarray}
\label{eq:E_perp}
E_\perp & = & \int_{4\pi} \totd \Omega\, \int_{r_{\rm in}}^{r_s} r^2 \totd r\,
                           \frac{1}{2} \rho v_\perp^2\\
\label{eq:E_r}
E_r     & = & \int_{4\pi} \totd \Omega\, \int_{r_{\rm in}}^{r_s} r^2 \totd r\, 
                          \frac{1}{2} \rho \left( v_r^2 - \langle v_r\rangle^2 \right),
\end{eqnarray}
where $v_\perp$ is any non-radial velocity component ($\theta$ and $\phi$ when
defined relative to the $z$ axis, or \emph{poloidal} and \emph{toroidal} otherwise),
the mean radial velocity $\langle v_r\rangle$ is computed as an instantaneous 
angle-average (e.g., \citealt{hanke2013}), and the inner radius for integration $r_{\rm in}$ is 
the gain radius\footnote{When comparing models with and without heating (\S\ref{s:non_exploding}), 
the inner radius for integration is taken to be that where the angle-averaged sound speed 
is maximal; c.f. Paper I.}. 

\begin{figure}
\includegraphics*[width=\columnwidth]{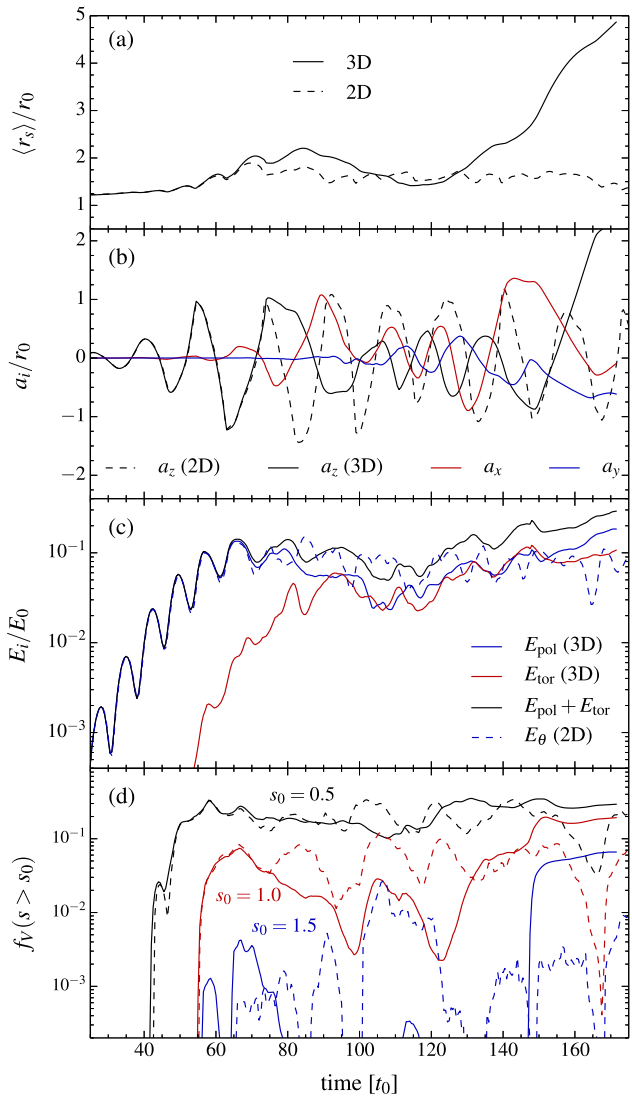}
\caption{Comparison between two SASI-dominated models that differ only in
their dimensionality: S08L1 (3D, marginally exploding) and s2d-08L1 (2D, non-exploding).
In both cases, an $\ell=1$ sloshing mode is initially excited; for the 3D model, the
sloshing axis is along $\theta=\phi=45^\circ$.
\emph{Panel (a):} Average shock radius. \emph{Panel (b):} Real spherical harmonic
coefficients (eq.~[\ref{eq:a_i}]) defined relative to the
initial sloshing axis in 3D or along the symmetry axis in 2D.
\emph{Panel (c):} Non-radial kinetic energies. In 3D, the poloidal and toroidal
directions are defined relative to the initial sloshing axis.
\emph{Panel (d):} Fraction of the post-shock volume that has
entropy higher than a given value $s_0$ (eq.~[\ref{eq:entropy_definition}]), as labeled.
The solid and dashed lines correspond to the 3D and 2D, models, respectively.}
\label{f:sfrac_B08_dim}
\end{figure}

Figure~\ref{f:sfrac_B08_dim} shows the time-dependent diagnostics
for the 2D and 3D comparison models (s2d-08L1 
and S08L1, respectively). The evolution of the angle-averaged shock radius is identical
up to time $t=60t_0$, 
showing the growth and saturation of the sloshing mode that is initially excited.
At this time, a sloshing mode in an orthogonal direction ($a_x$) has achieved
a noticeable amplitude.
Thereafter, the average shock
radius undergoes a larger shock excursion in 3D relative
to 2D. Nonetheless, this initial expansion dies down over
a time equivalent to several SASI oscillations. At time $t=115t_0$
a new expansion begins, and the 3D model
transitions to a runaway.
Meanwhile, the 2D model does not show any significant increase
in its average shock radius, achieving a statistical steady-state.

The spherical harmonic coefficients indicate that the $z$
component, initially excited, achieves the same saturation
amplitude in 2D and 3D. This is expected from estimates 
of the effect of parasitic instabilities 
\citep{guilet2010} and also manifests in the SASI-only tests 
in Appendix~\ref{s:code_tests}.
The final phase of spiral mode growth that precedes explosion 
in 3D is characterized by a comparable amplitude in all three
(out-of-phase) real spherical harmonic coefficients. 
Interestingly, in this second expansion phase 
none of the orthogonal sloshing components
in 3D achieves a higher saturation amplitude 
than the 2D mode except after runaway expansion has set in. 

\begin{figure}
\includegraphics*[width=\columnwidth]{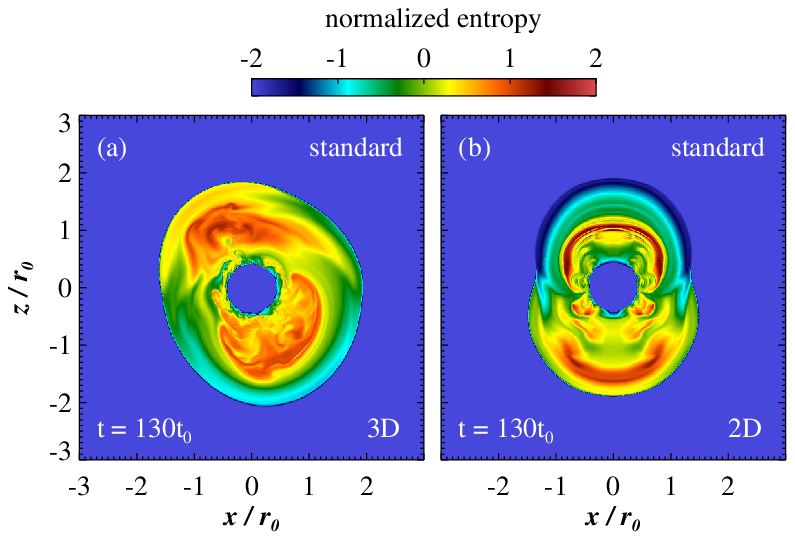}
\includegraphics*[width=\columnwidth]{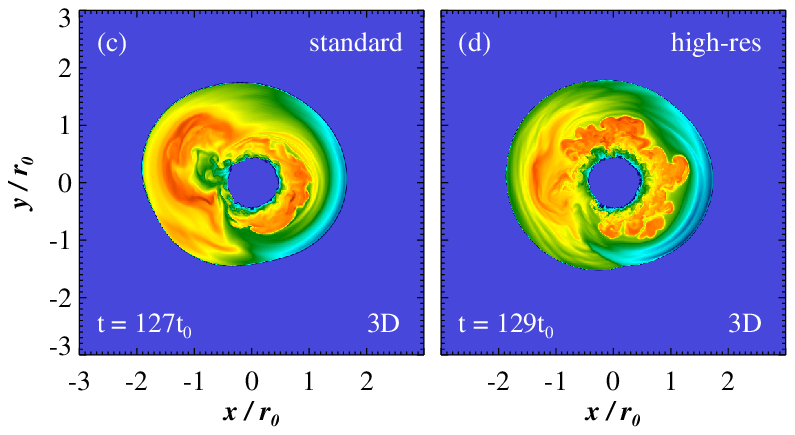}
\caption{Entropy slices in SASI-dominated models with
the same parameters and initial conditions except dimensionality and/or resolution,
as labeled.
\emph{Panels (a-b):} Slices on the $xz$ plane for 3D and 2D models. In the former,
the plane is defined relative to the sloshing mode that is initially excited
($\theta=\phi=45^\circ$). While the 2D model reaches higher entropy, the sizes
of the bubbles are smaller than in 3D. \emph{Panels(c-d):} Slices on the $xy$ plane in 3D
models with different resolution. The plane is perpendicular to the sloshing axis. 
The model with high angular resolution displays a larger degree of small-scale turbulence. 
Compare with Figs.~\ref{f:sfrac_B08_dim} and \ref{f:sfrac_B08_res}.}
\label{f:ents_slices_dimres}
\end{figure}

\begin{figure}
\includegraphics*[width=\columnwidth]{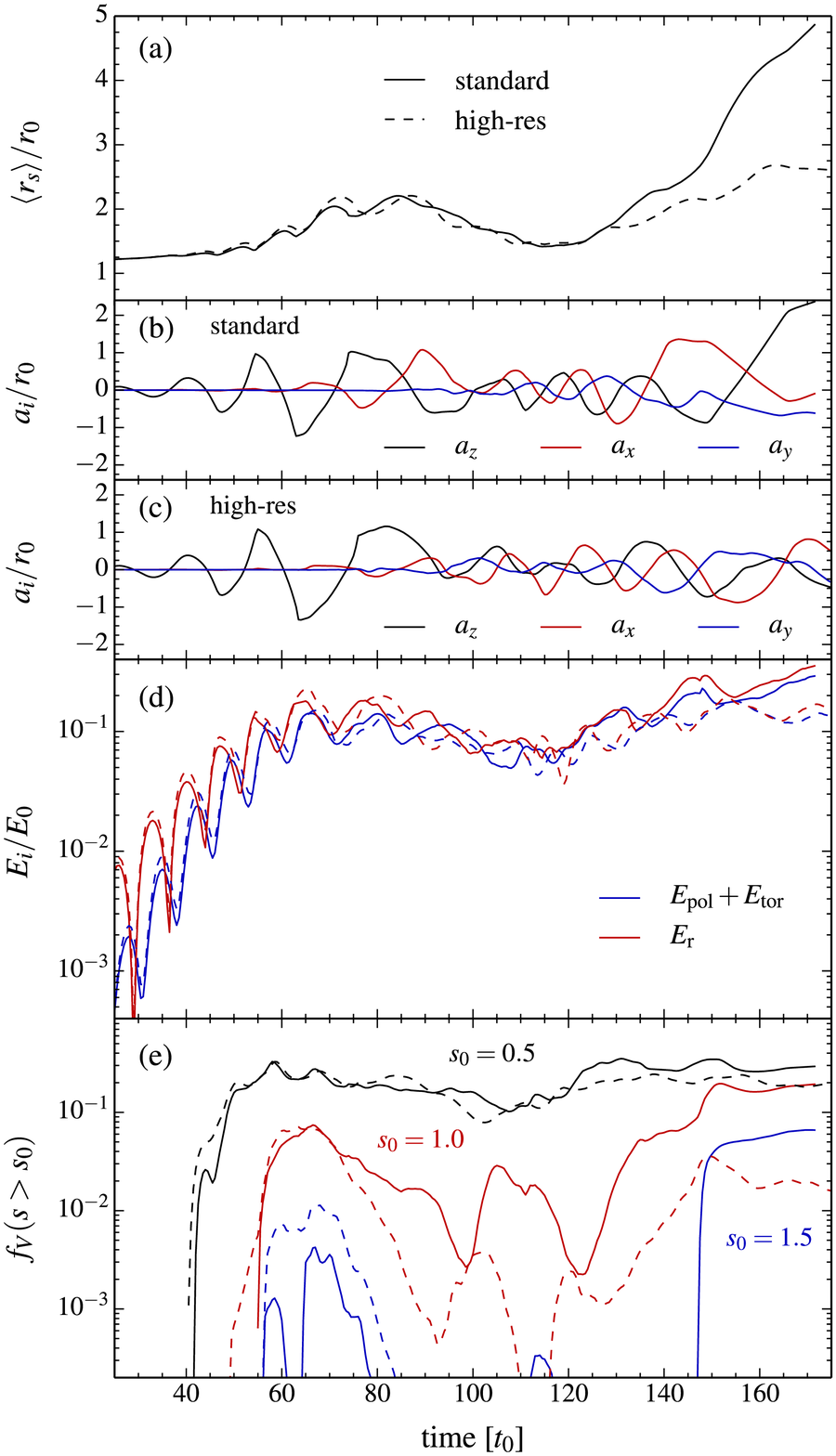}
\caption{Comparison between two SASI-dominated
models that differ only in their angular resolution: S08L1 (standard resolution,
marginally exploding) and S08L1-hr (high resolution, non-explodig). 
In both cases, a sloshing mode along $\theta=\phi=45^\circ$ is initially excited.
\emph{Panel (a):} Average shock radius. \emph{Panels (b) and (c):} Real spherical harmonic
coefficients (eq.~\ref{eq:a_i}) defined relative to the initial sloshing
axis. \emph{Panel(d):} Transverse and radial kinetic energies (eq.~[\ref{eq:E_perp}] and \ref{eq:E_r}). 
\emph{Panel (e):} Fraction of the post-shock volume that has
entropy higher than a given value $s_0$ (eq.~[\ref{eq:fv_definition}]), as labeled,
with the entropy is defined as in equation~(\ref{eq:entropy_definition}).
Solid and dashed lines correspond to standard and high resolution, respectively.}
\label{f:sfrac_B08_res}
\end{figure}

Also shown in Figure~\ref{f:sfrac_B08_dim} are the
non-radial kinetic energies in the gain region. Initially, the poloidal
kinetic energy in 3D (defined relative to the sloshing
axis excited) is identical to the $\theta$ kinetic energy
in 2D. The energies begin to evolve differently once the $a_x$ and $a_y$ modes are excited, 
manifesting as an increase in the toroidal kinetic energy. This toroidal
energy decreases
as the poloidal energy saturates, at the time when
the shock is retracting. The sum of poloidal and toroidal
energies remains comparable to the $\theta$ energy in 2D.
Once the spiral mode achieves noticeable amplitude
at $t\sim 100t_0$, both
toroidal and poloidal energies increase monotonically.
The time at which the final expansion of the shock
begins coincides with the time
when the sum of the non-radial kinetic
energies in 3D exceeds that in 2D. By the end of the
exploding simulation, this difference is a factor
of several.

The volume fraction diagnostic shows that in fact
the 2D model achieves larger overall fractions of the
volume with high entropy than the 3D model. This would
normally indicate that the 2D model achieves more favorable
conditions for explosion. Upon closer inspection, however, we 
find that while the 2D model contains structures with
higher entropy than in the 3D model, these structures
occupy disconnected volumes, as shown by the
$xz$ entropy slices in Figure~\ref{f:ents_slices_dimres}a-b.
These structures also live for a shorter time, as indicated
by the oscillatory pattern in the $f_V$ diagnostic
for the 2D model. Thus, achieving explosion conditions is 
not only related to
the size of the bubbles or the magnitude of the 
entropy in these structures, but also involves the time during
which these structures can survive.

Regarding the effect of resolution, Figure~\ref{f:sfrac_B08_res} shows 
the diagnostic quantities for the two 3D models that
differ only in their angular resolution, one exploding (S08L1)
and the other non-exploding (S08L1-hr; c.f. Figure~\ref{f:shock_critical_resolution}). 
The evolution of both models is qualitatively very similar until
$t\simeq 130t_0$, after which the evolution bifurcates. 
The spherical harmonic coefficients show that while 
the evolution of the sloshing mode that was initially 
excited is very similar in both cases, the
amplitude of $a_x$ 
is larger in the lower resolution model at the time
when explosion begins. In other
words, a stronger spiral SASI mode can be the
difference between explosion and failure.

This subtle difference is also reflected in the kinetic energies
in the gain region.
Both radial and non-radial energies are initially
larger in the higher resolution model, as expected
from the known dependence of the SASI growth rates with
angular resolution \citep{FT09a}. The energies 
in the two models
remain close to each other during the period of shock retraction,
and bifurcate only 
around the time when runaway sets in.

The volume fraction diagnostic $f_V$ is more informative
in this case, indicating that
the lower resolution model is able to maintain
a larger volume with high-entropy
relative to the high-resolution model 
after time $t\simeq 70t_0$.

Figure~\ref{f:ents_slices_dimres} also shows two snapshots
of the entropy on the $xy$ plane (defined relative to the
initially excited sloshing mode) at similar stages
prior to the bifurcation in the
evolution of these two models. The snapshots suggest
that the smaller scale turbulence in the higher
resolution model is related to
more efficient fragmentation of high-entropy bubbles. We interpret
the failure to explode in model S08L1-hr as
due to the shorter life of these
high-entropy bubbles given the more efficient disruption by
turbulence. 
This interpretation is a conjecture at present, requiring
further study to be established as the definitive cause for
the failure of higher resolution models.
An interesting question is whether this shredding of 
bubbles can be characterized in terms
of parasitic instabilities as 
for SASI modes (e.g., \citealt{guilet2010}).

\subsection{Spiral mode with heating: large shock excursions}
\label{s:non_exploding}

\begin{figure}
\includegraphics*[width=\columnwidth]{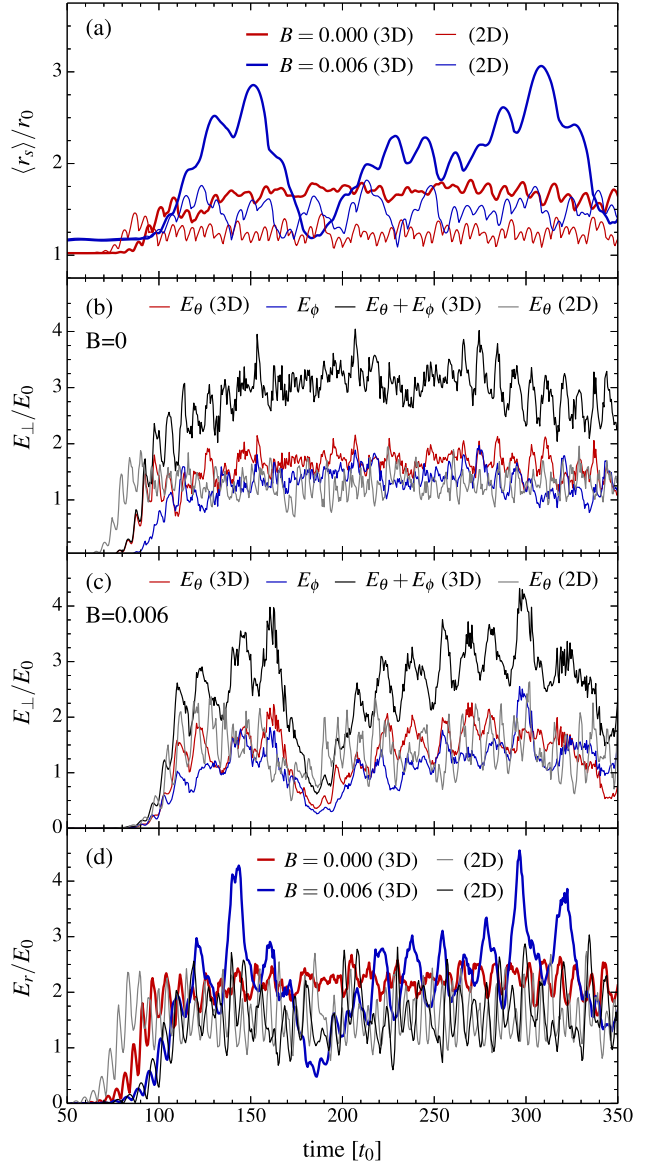}
\caption{Relation between shock expansion and kinetic energy in selected 
non-exploding SASI-dominated models, illustrating the effect of neutrino
heating and dimensionality (models s2d-00, s2d-06, S00dv, S06dv; see Table~\ref{t:models}). 
\emph{Panel (a):} Average shock radius. \emph{Panel (b) and (c):} Non-radial kinetic
energy (eq.~[\ref{eq:E_perp}]) for models with B=0 and B=0.006, respectively. \emph{Panel (d):} Radial
kinetic energy in the fluctuating component of the flow (eq.~[\ref{eq:E_r}]). The 
operation of spiral modes in 3D leads to a larger amount of non-radial
kinetic energy than in 2D. Episodic shock excursions are related to the formation
of large-scale high-entropy bubbles when neutrino
heating is included. When the bubbles are shredded and an explosion
fails to start, the shock retreats and the cycle resets.}
\label{f:noexpl_energy}
\end{figure}

Another interesting result of this investigation is that
SASI-dominated models in 3D that do not explode, but which are close
to the critical neutrino luminosity, display shock excursions that
are much larger than in their 2D counterparts. This effect is visible
in Figure~\ref{f:shock_critical} for the model with $B=0.007$.
Here we investigate the origin of this effect, and discuss the
potential for a larger difference between 2D and 3D
once a realistic EOS, with alpha particle recombination, is included in 
the evolution.

For this analysis, we carry out two additional non-exploding models,
each in 2D and 3D: one with no heating (s2d-00 and S00dv, respectively)
and another with $B=0.006$ (s2d-06 and S06dv, respectively).
Figure~\ref{f:noexpl_energy}a compares the evolution of the average shock radius for 
these four models. Two features stand out. First, the 2D models have systematically 
smaller shock radii than their 3D counterparts. Second, models with heating undergo
episodes in which the shock executes large excursions. In the 2D model with heating, 
these excursions last for a shorter time and have smaller amplitudes than in 3D.

Inspection of the 3D model with heating shows that these excursions
are associated with the appearance of large-scale, high-entropy
bubbles. An expansion of the
shock against a supersonic incident flow leads to an increase in
the Mach number in the frame of the shock, leading to a larger
entropy jump and hence to the formation of a high-entropy bubble
with a size comparable to the portion of the shock that is expanding 
(e.g., Paper I). As is the case in 2D, the formation of
large enough bubble leads to 
the interruption of the SASI cycle. If 
a runaway expansion fails to ensue, the bubble is shredded, either by
turbulence or by external bulk motions (e.g., downflows),
and the shock retracts. A larger shock radius 
means a larger volume subject to neutrino heating where high-entropy bubbles can grow. 

Figure~\ref{f:noexpl_energy}b-c also shows the non-radial
kinetic energies for the four comparison systems (c.f. eq.~[\ref{eq:E_perp}]).
Given that models without heating are included in the comparison, 
the lower radius for computation of these energies is 
that where the angle-averaged sound speed is maximal (e.g., Paper I). 
The resulting (total) non-radial kinetic energies are larger in 3D than in 2D, 
whether neutrino heating is included or not. It is not surprising
then that the average shock radius is larger in 3D than in 2D, as 
more kinetic energy inside the shock provides additional 
support against the ram pressure of the collapsing stellar 
core (e.g., \citealt{murphy2012,couch2015a}).

For models without heating, in which non-radial
motions are solely driven by the SASI, the non-radial kinetic
energies in the fully non-linear phase satisfy 
$E_\theta^{3D} \sim E_\phi \sim E_\theta^{2D}$.
Hence, the spiral modes of the SASI are able to approximately double the total
non-radial kinetic energy in the post-shock domain. Inspection of
Figure~\ref{f:sfrac_B08_dim}b as well as tests in 
Appendix~\ref{s:code_tests} indicate that the saturation
amplitude of individual sloshing modes, quantified by the real spherical
harmonic coefficients (eq.~[\ref{eq:real_spherical_harmonics}]), is very
similar in 2D and 3D (this is also a prediction from the analysis of \citealt{guilet2010}). 
Recall that two or three linearly-independent
real spherical harmonic coefficients oscillating in phase are indicative of a single
sloshing mode, thus their individual amplitudes will not reach the maximum
possible value (their root-mean-square sum will). A spiral mode appears to be able 
to tap into the full amplitude of its individual 
constituent (orthogonal) sloshing modes, hence roughly doubling the kinetic energy. 

For the cases with heating, the non-radial kinetic energy in 2D is slightly larger
than the individual $\theta$ or $\phi$ energies in 3D, but still smaller than their sum.
The episodic shock excursions in 3D are concurrent with increases
in the non-radial kinetic energy. However, the overall magnitude of
the non-radial 3D energies are very close to those of the model  
without heating.

The key feature that sets apart the model with large shock excursions 
from others is the magnitude of the fluctuating radial kinetic energy 
(eq.~[\ref{eq:E_r}]), as shown in Figure~\ref{f:noexpl_energy}d. 
Whereas in 2D the radial kinetic energy is comparable to that in the 
$\theta$ direction whether heating is included or not, in 3D the 
total non-radial kinetic energy exceeds its radial counterpart by 
about $\sim 50\%$ during most of the time. However, during episodes of 
transient shock expansion, the radial kinetic energy of the model with $B=0.006$
shows large increases that match or even exceed
the total non-radial kinetic energy.

We interpret this increase in radial kinetic energy as being
a consequence of the large high-entropy bubbles having internal
structure and being convective, as can be seen from 
Figure~\ref{f:entropy_slices_explosion}. In a fluid where buoyant
convection operates, the relation $E_{r} \simeq E_\theta + E_\phi$
holds \citep{murphy2012}. The fact that the radial kinetic energy
remains large only for a short period of time is related to the
short lifetime of the large-scale bubbles, which are shredded 
prior to shock retraction.

The large shock excursions shown by the 3D SASI-dominated models
with neutrino luminosities close to critical hold promise
for further reductions in this critical heating rate once
a more realistic EOS is included.
If the effect of alpha particle recombination is included,
the ratio of nuclear dissociation energy to gravitational potential
energy at the shock does not increase with radius 
(e.g. \citealt{FT09b}), and hence the system loses less energy
upon expansion. A characteristic radius $r_\alpha$ where this effect
becomes dynamically important is that were the nuclear binding energy
of alpha particles equals the gravitational binding energy,
\begin{equation}
r_\alpha \simeq 254 M_{1.3}\textrm{ km}.
\end{equation}
Taking $r_0 = 100$~km for a characteristic stalled shock radius
without neutrino heating, shock excursions such as 
those of model S06dv should very likely
lead to a runaway expansion. 
This conjecture will be tested in future work.


\section{Summary and Discussion}
\label{s:summary}

We have investigated the properties of SASI- and convection-dominated
core-collapse supernova explosions in three dimensions using 
parameterized hydrodynamic simulations. This approach allows to 
isolate the effect of hydrodynamics instabilities on the explosion 
mechanism from uncertainties in progenitor models, the EOS of dense matter, 
or an incomplete treatment of neutrino effects. By generating
sequences of models well within the parameter regimes where
either the SASI or convection dominates, we characterized
the effect of adding a 3rd spatial dimension on the dynamics.
Our main results are the following:
\newline

\noindent
1. -- SASI-dominated explosions do exist in 3D, and their evolution
      is qualitatively different from convection-dominated explosions.
      (Fig.~\ref{f:entropy_slices_explosion}). For systems in which both
      the mass accretion rate and protoneutron star radius change slowly relative
      to the thermal time in the gain region, the value of the convection parameter 
      $\chi$ (eq.~[\ref{eq:chi_definition}]) at the time when the shock stalls is a 
      good predictor of the explosion path to be taken by the system, if the 
      neutrino luminosity is high enough.
\newline

\noindent
2. -- SASI-dominated systems
      can explode with a lower neutrino luminosity in 3D than in 2D 
      (Fig.~[\ref{f:shock_critical}], Table~\ref{t:models}). This difference
      is related to 
      the ability of spiral modes to generate more non-radial
      kinetic energy than a single sloshing mode, resulting in a larger
      average shock radius and hence generating more favorable conditions
      for the formation of large-scale, high-entropy bubbles (Fig.~\ref{f:sfrac_B08_dim}).
      For the baseline angular resolution employed here ($\Delta\theta \simeq \Delta\phi\simeq 2^\circ$
      at the equator), the critical neutrino luminosity in 3D is $\sim 20\%$ lower than
      in 2D.
\newline

\noindent
3. -- Convection-dominated explosions 
      show a much smaller
      change in the critical heating rate with
      increasing dimensionality. In fact, no significant difference was found
      between 2D and 3D for the spacing in heating rate explored here
      ($\sim 8\%$). This result is consistent with previous studies (e.g., \citealt{hanke2012}). 
      Doubling the angular resolution of the marginally
      exploding models yields earlier explosions in 2D and minor
      changes in explosion time for 3D (Table~\ref{t:models}). 
\newline

\noindent
4. -- Doubling the angular resolution in $\theta$ and $\phi$ decreases the difference
      in critical neutrino luminosity between 2D and 3D for SASI-dominated 
      explosions (Figs.~\ref{f:shock_critical_resolution} and 
      \ref{f:sfrac_B08_res}). 
      We interpret this reduction as a consequence of the
      higher efficiency of 3D turbulence at higher resolution for disrupting
      the large-scale, high-entropy bubbles needed to launch
      an explosion (Fig.~\ref{f:ents_slices_dimres}). The exact
      dependence of this increase in the critical neutrino luminosity
      with resolution in 3D was not obtained, but based on the healthy
      explosion obtained in a 3D model with $10\%$ lower heating rate
      than the marginal 2D model ($B=0.009$, both at high resolution; Table~\ref{t:models}), 
      we infer that this difference is shallower than 
      $B_{\rm crit}^{3D}\propto (\Delta\theta\Delta\phi)^{-0.085}$.
      This sensitivity to resolution also highlights the need for
	a better understanding of the systematic uncertainties associated
	with the use of finite volume methods to model supernova flows, 
	in which the spatial resolution is limited by current computational 
	resources (e.g., \citealt{porter94,radice2015}).
\newline

\noindent
5. -- Non-exploding SASI-dominated models have larger average
      shock radii in 3D than in 2D, which we interpret to be
      a consequence of the higher non-radial
      kinetic energy generated by spiral modes (Fig.~\ref{f:noexpl_energy}).
      While the saturation of individual sloshing modes appears to be
      very similar in 2D and 3D (Fig.~\ref{f:sfrac_B08_dim}, Appendix~\ref{s:code_tests}), 
      a spiral mode can generate more kinetic energy
      than a single sloshing mode (e.g., Figure~\ref{f:shock_energies_L1z}). 
      This result is in agreement with the
      findings of \citet{hanke2013}, and in 
      conflict with the interpretation of \citet{iwakami08} that saturation
      in 3D is smaller because the same amount of kinetic energy is shared between a larger number of modes.
      The exact limiting factor to the energy available for a spiral mode 
      was not investigated here, but it is worth pursuing in future studies.
\newline

\noindent
6. -- Very large shock excursions can result in 3D when the heating
      rate is close to (but lower than) the critical value for an
      explosion in SASI-dominated models. These excursions 
      are simultaneous with the appearance of
      high-entropy bubbles
      generated by the expanding shock, and are smaller in magnitude
      (by a factor $\sim 2$) in 2D. The non-radial kinetic energy is
      generally larger than the radial kinetic energy when spiral modes
      operate, except when large bubbles are formed and the shock expands,
      in which case the relation $E_r \sim E_\theta+E_\phi$ holds.
      We conjecture that this is due to the internal structure of bubbles,
      which is likely convective. The large shock excursions open the
      possibility of further improvements in explosion conditions in 3D
      when the effect of alpha particle recombination is included.
\newline

While the behavior of the $27M_\sun$ model of \citet{hanke2013}
during the time when spiral modes operate appears to be consistent with our results, 
the period of spiral-SASI-driven shock expansion is too short to 
make an unambiguous parallel. 
The maximum amplitudes of the sloshing modes involved and the total
transverse kinetic energies exceeded the values obtained in 2D for a
brief period. Unfortunately, just as the shock begins
to turn around and grow at a rate faster than in 2D, the Si/SiO composition 
interface is accreted through
the shock, triggering a large expansion that makes the model convective
again. 

The relation between growth of the non-radial kinetic energy
and approach to explosion was noted by \citet{murphy11} and 
\citet{hanke2012}. Intuitively, one expects accreted
material to spend more time in the gain region when
the flow is non-radial and the gain region is larger due
to the larger shock radius (e.g., \citealt{murphy08}).
An interesting way to think about the reduction in 
critical neutrino luminosity when going from 1D to 2D/3D
is that due to the non-laminar stresses, a given shock radius
requires less thermal pressure to be maintained and thus 
less neutrino heating to drive an expansion from this position
than if the flow were laminar \citep{couch2015a}. This is 
equivalent to saying that, in analogy with accretion disks, 
the effective gravity felt by the fluid
inside the shock is smaller due to the action of bulk or 
turbulent stresses, 
\begin{equation}
g_{\rm eff} \sim \frac{GM(r)}{r^2} - \frac{v^2_{\rm fluct}}{r},
\end{equation}
where $v^2_{\rm fluct} = \langle v_r^2 - \langle v_r\rangle^2 
+ v_\theta^2+v_\phi^2\rangle$. This is also consistent
with the critical r.m.s. Mach number threshold proposed
by \citet{mueller2015}. Regardless of the conceptual framework
used to interpret it, the improved ability of spiral modes to
extract non-radial kinetic energy from the accretion flow
appears to
improve conditions for explosion.

While we have found a region of parameter space
in which 3D is more favorable to explosion than
2D, it is fair to ask whether this parameter space
is actually realized in Nature. The $27M_\sun$ 
model of \citet{hanke2013} is genuinely SASI-dominated
until the Si/SiO interface is accreted, as diagnosed
by the evolution of $\chi$, which lingers around $2$
before the composition interface falls through the shock. 
In contrast, the same progenitor evolved by \citet{couch2014} 
(using $f_{\rm heat}=1.00$) yields $\chi$ closer to $3$ when
the shock stalls, showing that the treatment of neutrino 
transport can be influential in determining the evolutionary path 
taken by the system. More generally, our knowledge of 
two key external ingredients to the supernova mechanism is
by no means complete. First, progenitor stars are likely to
be members of interacting binary systems (e.g., \citealt{sana2012,smith2014}),
intrinsically multidimensional (e.g., \citealt{arnett2011,couch2015b})
and with a degree of rotation yet to be convincingly 
established (e.g., \citealt{fuller2015}). Second, the EOS
of matter at supra-nuclear densities remains uncertain despite
significant recent progress (e.g., \citealt{lattimer2012}).

In the end it may well turn out 
to be that, as suggested by \citet{abdikamalov2014}, the SASI is 
relevant mostly for high accretion rate progenitors that 
fail to explode. Because neutron star formation generally precedes
collapse to a black hole (e.g., \citealt{oconnor2011}), the
same predictions for the neutrino and gravitational
wave emission for systems with strong SASI activity 
(e.g., \citealt{mueller2013,tamborra2014}) should apply. In fact, 
the higher accretion rates in failed systems are expected 
to yield a more intense neutrino signal than exploding 
models \citep{oconnor2013}. This should improve prospects for identification
of a Galactic event even if dedicated surveys (e.g., \citealt{kochanek2008}) fail to detect
the electromagnetic signature predicted for these 
systems \citep{nadezhin1980,lovegrove2013,piro2013,kashiyama2015}.

If on the other hand progenitor properties turn out to
favor SASI-dominated explosions once uncertainties have
been resolved, the effect of
pre-collapse perturbations on the development of a
spiral SASI mode remains as a potential obstacle.  The advective-acoustic
cycle involves the interplay of coherent perturbations that
require a relatively smooth background flow to develop. The
work of \citet{guilet2010} indicates that large amplitude
perturbations can disrupt the acoustic feedback in the cycle
and lead to ineffective growth. Furthermore, the 
work of \citet{mueller2015} suggests that if physically-motivated
pre-collapse perturbations are used, the shock acquires
a nearly static asphericity, with detrimental conditions
for the development of the SASI (but providing an alternative
channel for the generation of non-radial kinetic energy).

Our results prove the principle that SASI-dominated explosions
can provide better conditions for
explosion in 3D than in 2D. For the effect to be considered
robust, however, a key set of physical ingredients needs to
be added. Future work will address the response of these types
of explosion to the inclusion of a realistic
EOS (including nuclear recombination), rotation in the 
accretion flow, the feedback from the
accretion luminosity on the gain region, and the time-dependence
of the accretion rate and neutron star radius. 
These studies will help clarify issues such as 
the behavior of the $20M_\sun$ model with sophisticated neutrino transport
reported in 
\citet{tamborra2014}, which in 3D undergoes an extended phase of SASI 
activity but fails to explode, while succeeding in 2D (T. Janka, private
communication). 

A more realistic (but still parametric) study would also
allow exploration of issues such as the elementary processes
that mediate the transition to explosion 
(e.g., \citealt{pejcha2012,F12,mueller2015}), the amount of angular 
momentum imparted to the neutron star by spiral 
modes, which depends on the mass cut at the time of explosion 
(e.g., \citealt{guilet2014}) as well as the potential
for spiral modes to amplify magnetic fields (e.g., \citealt{endeve2012})
in an explosion context.
The results of such studies would inform the analysis of more sophisticated
models, in which the high computational cost precludes 
a systematic exploration of parameter space.

Ultimately, the magnitude of the improvement in efficiency on the
neutrino mechanism introduced by the SASI in 3D, if all conditions are 
favorable, is of the order of $\sim 10\%$. While this contribution
is relatively modest, it can help to tilt systems that would 
be otherwise marginal towards a robust explosion (as is the case when, e.g.,
the neutrino cross sections are modified at the $\sim 10\%$ level in the
right direction; \citealt{melson2015b}).

\section*{Acknowledgments}

I thank Dan Kasen, Eliot Quataert, Thomas Janka, Thierry Foglizzo, J\'er\^ome Guilet,
Ralph Hix, Sean Couch, and John Blondin
for helpful discussions and/or comments on the manuscript.
The anonymous referee provided constructive comments that improved the presentation
of the paper.
RF acknowledges support from the University of California Office of the President, and
from NSF grant AST-1206097.
The software used in this work was in part developed by the DOE-supported ASC / 
Alliance Center for Astrophysical Thermonuclear Flashes at the University of Chicago.
This research used resources of the National Energy Research Scientific Computing
Center (NERSC), which is supported by the Office of Science of the U.S. Department of Energy
under Contract No. DE-AC02-05CH11231. Computations were carried out at
\emph{Carver} and \emph{Hopper} (repo m2058).
\newline

\appendix

\section{Extension of the split PPM solver in {\textsc FLASH3.2} to 3D spherical coordinates}
\label{s:code_tests}

Here we describe the extension to 3D spherical polar coordinates of the split PPM solver present 
in the public version of {\textsc FLASH3.2}. We then
describe tests of this modified code. The test cases can be used to diagnose any 
hydrodynamics code in 3D spherical coordinates.

\subsection{Implementation}

The split PPM solver embedded in {\textsc FLASH3.2} is based on 
{\textsc PROMETHEUS} \citep{fryxell1989}, which implements the method of 
\citet{colella84}. The public FLASH version
requires only a few modifications to run on a grid defined in 3D spherical 
coordinates. The modifications are:
\begin{enumerate}
\item Allowing 3D spherical geometry in the subroutine {\tt Hydro\_detectShock},
      and computing the appropriate velocity divergence;
\item Modifying the ratio of time step to cell spacing $\totd t / \totd x$
      in the subroutine {\tt hydro\_1d} to account for the proper cell spacing in the
      azimuthal direction, $\totd x_{\phi} = r\sin\theta \totd \phi$
\item Implementing a safeguard in the subroutine {\tt avisco}, whenever division by 
      $\sin\theta$ leads to singular behavior in the velocity divergence.
      This is the only subroutine where this divisor is evaluated at the cell
      face, where it can vanish if the grid extends to the polar axis. In practice,
      a small floor value of $\sin\theta$  ($\sim 10^{-6}$) can be imposed.
\item Modifying the arguments to the subroutine {\tt Hydro\_detectShock}, called
      by {\tt hy\_ppm\_sweep} in order to use the correct coordinates.
\end{enumerate}

In addition to these basic modifications, we have experimented with 
two additional changes that can improve the accuracy of the code:
\begin{enumerate}
\item The fictitious accelerations that account for the centrifugal and
      Coriolis forces in curvilinear coordinates are computed with
      velocities evaluated at the beginning of the time-step. For 
      consistency with PPM, these accelerations should be 
      time-centered. We implement this modification by evaluating
      $1/2$ of the acceleration forward in time after the PPM update.
      This is straightforward
      because the force along a given direction depends 
      linearly on the corresponding velocity component.
      Tests indicate that this modification leads to minor differences
      in the results.
 
\item The default reflecting boundary condition in the
      $\theta$ direction at the axis
      can be improved by making use of the data across the axis
      to fill the ghost cells.
      For example, for the first active cell
      next to the axis (with center coordinate $\theta = \theta_1$), the contents of its neighboring
      ghost cell can be set to satisfy
      \begin{equation}
      \label{eq:transmit_bndcnd}
      A(r,-\theta_1,\phi) = A(r,\theta_1,\phi + \pi),
      \end{equation}
      where $A$ is an arbitrary variable. The sign of the transverse velocities
      $v_\theta$ and $v_\phi$ must be changed to account for 
      the sign change in the corresponding unit vectors across the axis.
      This type of operation requires ghost cell exchange between processors
      that contain cells located at opposite sides of the axis.
 
\end{enumerate}

Additional changes are needed only when using a non-uniform grid.  These
changes involve generating the non-uniform coordinates, using the exact cell
volumes and areas (instead of linearized approximations), and passing the
correct vector of cell spacings to the PPM routines instead of a constant
value. These modifications are further described in \citet{F12}.

\subsection{Tests}

\begin{table}
\centering
\begin{minipage}{8cm}
\caption{Test models evolved. Columns from left to right show model name,
type of initial perturbation applied, use of a hybrid Riemann solver
at shocks, type of boundary condition used at the polar axis
in the $\theta$ direction, and root-mean-square amplitude of the $\ell=1$
spherical harmonic coefficient along the excited dipole axis (for reference,
the 2D value is $\Delta a_1/r_0 = 0.544$). The initial 
perturbation has the form of
an overdense shell with angular dependence given by a real spherical 
harmonic: $Y_z$, $Y_x$, or $Y_d \equiv Y_z/\sqrt{2} + (Y_x+Y_y)/2$. When
no explicit perturbation is applied, initial transients result in a spherically-symmetric
perturbation. For comparison, 1D and 2D versions of each model are also 
evolved.\label{t:test_models}}
\begin{tabular}{lcccc}
\hline
{Model} &
{Pert.} &
{Hybrid?} &
{Axis Bnd.} &
{$\Delta a_1/r_0$}\\
\hline
T-L0-hyb  & none     & yes & reflect & ...\\
T-L0-std  &          & no  &         & ...\\
\noalign{\smallskip} 
T-L1z-ref & $Y_z$    & yes & reflect & 0.534 \\
T-L1x-ref & $Y_x$    &     &         & 0.530 \\
T-L1d-ref & $Y_d$    &     &         & 0.544 \\
\noalign{\smallskip}
T-L1z-trm & $Y_z$    & yes & transmit & 0.533 \\
T-L1x-trm & $Y_x$    &     &          & 0.526\\
T-L1d-trm & $Y_d$    &     &          & 0.552\\
\hline
\end{tabular}
\end{minipage}
\end{table}

We test the reliability of the hydrodynamic solver in 3D by using 
the parameterized accretion shock setup described in \S\ref{s:setup}
without neutrino heating. The system is then only unstable to the
SASI, with no convection. By exciting individual modes of system,
the numerical solution can be compared directly with predictions
from linear stability analysis \citep{FT09a}.

The parameters of the setup are an initial ratio of star to
shock radius $r_*/r_0 = 0.5$, adiabatic index $\gamma=4/3$,
no nuclear dissociation at the shock ($\varepsilon = 0$), and
the same parameterized neutrino source term in 
equation~(\ref{eq:neutrino_source}) but without
neutrino heating ($B = 0$). This is the same set of parameters
used in \citet{F10}, which has the advantage that
only the fundamental $\ell=1$ SASI mode is unstable. In the absence
of rotation, the modes are degenerate in $m$, serving as
a good diagnostic of the isotropy of the code. This is a
global problem involving subsonic flow, where a
delicate balance between pressure gradients and gravity
needs to be maintained. Any obvious errors in the code
manifest immediately.

The tests differ in the type of initial
perturbations applied and in the boundary condition
at the axis. SASI modes are excited by dropping an
overdense shell with a given angular dependence, as
described in \S\ref{s:setup}.
For comparison, 1D and 2D versions
are also evolved. We used the standard angular resolution
($\Delta\theta \simeq \Delta \phi \simeq 2^\circ$ at the equator, \S\ref{s:setup}) 
for all tests. Table~\ref{t:test_models} summarizes
the parameters of each test simulation. Model names start with
T for ``test'', and then indicate the type of initial perturbation
and boundary condition employed (e.g., T-L1z-ref corresponds to $\ell=1$
sloshing mode along the $z$ axis, with reflecting boundary condition
at the axis).

The tests probe four aspects of the code: (1) the ability
to maintain spherical symmetry and agreement with 1D and 2D
versions, (2) the ability to maintain axisymmetry and
its agreement with the 2D version, (3) the effect of
using a reflecting or `transmitting' boundary condition
at the axis, and (4) the isotropy of the code in 3D (symmetry
along an arbitrary axis).

\subsubsection{Spherical symmetry}

The first test simply lets the accretion flow  evolve
in the absence of perturbations. Initial transients
generate an $\ell=0$ perturbation that damps on a timescale
of several $100t_0$, because the corresponding SASI mode
is stable.

\begin{figure}
\includegraphics*[width=\columnwidth]{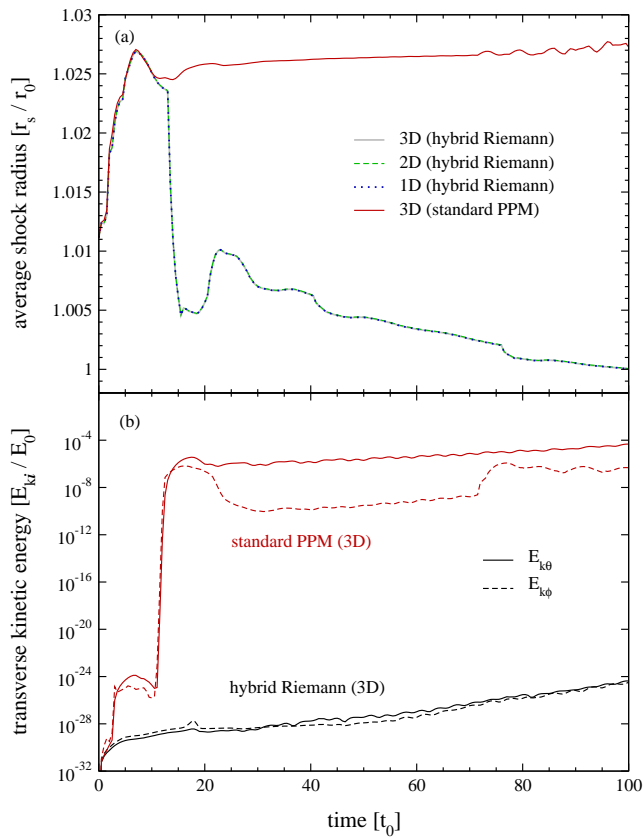}
\caption{Test of the ability of the code to maintain spherical symmetry 
in the absence of explicit perturbations. \emph{Panel (a)}: Average shock radius
as a function of time for 1D, 2D, and 3D models without initial perturbations.
Also shown is a 3D model that does not use the hybrid Riemann solver. 
\emph{Panel (b):} Transverse kinetic energies for 3D models without perturbations,
showing hybrid Riemann (black) and standard split PPM (red). Using the hybrid
Riemann solver keeps spurious non-radial velocity perturbations from growing too much above
the numerical noise.}
\label{f:spherical_hybrid_test}
\end{figure}

To achieve a clean test setup, it is important to damp
numerical perturbations that arise spontaneously. We find
that significant numerical noise arises from the shock
in 3D models if the standard PPM solver is used in all
cells. This is illustrated in Figure~\ref{f:spherical_hybrid_test}, which
shows the angle-averaged shock radius for models T-L0-std and
T-L0-hyb, as well as the magnitude of the non-radial kinetic
energies in the $\theta$ and $\phi$ directions. 

It is well known that a hybrid Riemann solver can eliminate
numerical problems at shocks aligned with the grid,
particularly the odd-even decoupling instability 
and carbuncle phenomenon \citep{quirk94}. {\textsc FLASH3.2}
allows the use of a HLLE solver inside shocks to increase
dissipation and damp these numerical instabilities.
Figure~\ref{f:spherical_hybrid_test} shows that when
this hybrid Riemann solver is used, spurious numerical 
perturbations (quantified by the non-radial kinetic
energies) remain close to the numerical noise, growing
slowly. 

Using the hybrid Riemann solver also enables excellent
agreement between 1D, 2D, and 3D.
This agreement is reached only above a certain radial resolution,
however, due to the steep density gradient at the base
of the flow, which results in the irregular misidentification
of shocks at low resolution. The baseline radial resolution adopted
in the study (a fractional radial spacing $\Delta r/r \simeq 0.45\%$ or
$0.26^\circ$) is the lowest resolution for which this agreement between
different dimensions is reached. 

\begin{figure}
\includegraphics*[width=\columnwidth]{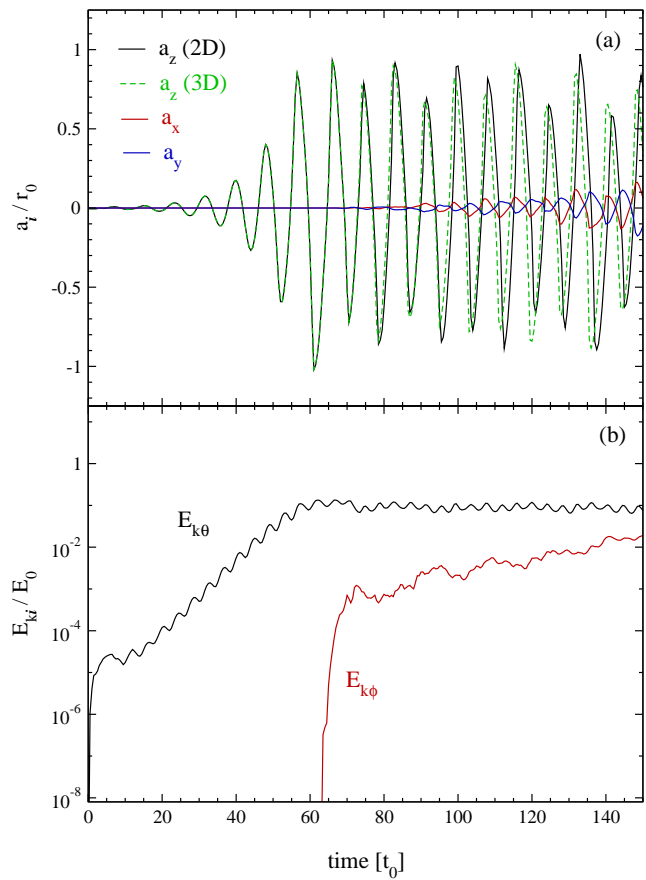}
\caption{Test of the ability of the code to maintain axisymmetry.
\emph{Panel (a)}: Real spherical harmonic coefficients (eq.~[\ref{eq:a_i}]) for model T-L1z-ref
and an identical run in 2D. In both cases an $\ell=1$ SASI sloshing mode
along the $z$ axis is initially excited.
The code remains axisymmetric until the SASI saturates. \emph{Panel (b):} Kinetic energies
in the polar ($\theta$) and azimuthal ($\phi$) directions for model T-L1z-ref. Upon
saturation of the SASI, the azimuthal kinetic energy grows rapidly from numerical
noise and triggers the growth of modes orthogonal to the $z$ axis.}
\label{f:shock_energies_L1z}
\end{figure}

\subsubsection{Symmetry around the $z$ axis}

The next test probes the ability of the code to maintain axisymmetry around the
$z$ axis. We evolve a 3D model for which an $\ell=1$ sloshing mode is excited
along the $z$ axis (T-L1z-ref). Given our choice of parameters, only the
fundamental $\ell=1$ mode should be unstable and therefore a clean sinusoidal
oscillation of the corresponding $\ell=1$ spherical harmonic coefficient 
(eq.~[\ref{eq:a_i}]) should be obtained.

Figure~\ref{f:shock_energies_L1z}a shows the evolution of the $z$ real spherical harmonic
coefficient (eq.~[\ref{eq:a_i}]) for model T-L1z-ref, and compares it with a 2D version
that excites an $\ell=1$ mode in the same way. The 3D model remains
axisymmetric up to the moment when the sloshing SASI mode saturates at $t\simeq
60t_0$. Thereafter, the azimuthal kinetic energy experiences rapid growth out
of numerical noise, as shown in Figure~\ref{f:shock_energies_L1z}b, leading to
the excitation of transverse sloshing modes in the $x$ and $y$ directions. This
growth in transverse motion originates in numerical noise at the polar axis.
Nonetheless, the sloshing mode along the $z$ direction remains nearly identical
between 2D and 3D up to a time $t\simeq 70t_0$, or $\sim 24,000$ time steps. 
At later times the primary difference lies in the phase of the oscillation.

The rms fluctuation in the $a_z$ coefficient between $t=75$ and $150t_0$ is
$\Delta a_1/r_0 = 0.544$ and $0.534$ in 2D and 3D, respectively, with a
relative difference of $2\%$ in favor of 2D. Also, note that the difference in
phase between the real spherical harmonic coefficients of the 3D 
model shows that a spiral mode is triggered.

\subsubsection{Axis boundary condition}
\label{s:axis_bnd}

\begin{figure}
\includegraphics*[width=\columnwidth]{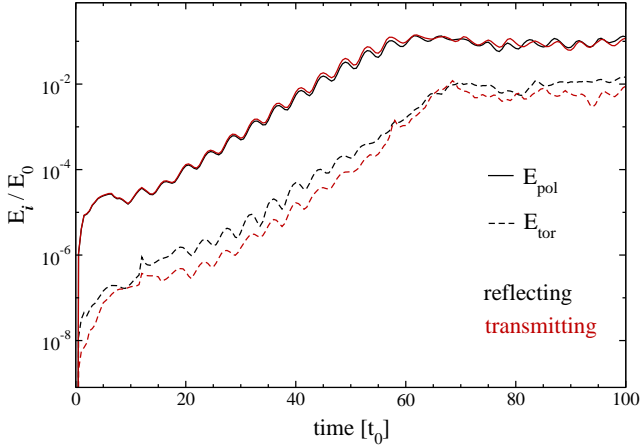}
\caption{Non-radial kinetic energies -- defined relative to the $x$-axis (eq.~\ref{eq:v_poloidal_x} 
and \ref{eq:v_toroidal_x}) --
as a function of time for models T-L1x-ref (black lines) and T-L1x-trm (red lines), in
which a sloshing SASI model along the $x$-axis is initially excited, and which use 
reflecting and transmitting boundary 
conditions in the $\theta$ direction at the axis, respectively. Solid lines
show poloidal and dotted lines toroidal kinetic energy.}
\label{f:energies_L1x}
\end{figure}

To obtain a quantitative measure of the improvement gained by using a
`transmitting' boundary condition in $\theta$ (eq.~\ref{eq:transmit_bndcnd}) instead of
a reflecting $z$ axis, we evolve two models in which an $\ell=1$ sloshing
mode along the $x$ axis is excited. 

As a diagnostic, we define poloidal and toroidal velocities relative 
to the $x$-axis:
\begin{eqnarray}
\label{eq:v_poloidal_x}
v_{\rm pol} & = &\frac{xy}{r\varrho}v_y + \frac{xz}{r\varrho}v_z - \frac{\varrho}{r}v_x\\
\label{eq:v_toroidal_x}
v_{\rm tor} & = & -\frac{z}{\varrho}v_y + \frac{y}{\varrho}v_z,
\end{eqnarray}
with $\varrho = \sqrt{y^2+z^2}$. These velocities are
the analog of $v_\theta$ and $v_\phi$, respectively.
In theory, the sloshing mode should remain symmetric around the $x$ axis, 
with the toroidal velocity $v_{\rm tor}$ serving as a diagnostic for numerical errors.

Figure~\ref{f:energies_L1x} shows the evolution of the poloidal and 
toroidal kinetic energies in models T-L1x-ref and T-L1x-trm, constructed with 
the corresponding velocities defined in equation~(\ref{eq:v_poloidal_x}) 
and (\ref{eq:v_toroidal_x}). In contrast to the sloshing mode along
the $z$ axis (Fig.~\ref{f:shock_energies_L1z}), for which the azimuthal
kinetic energy undergoes rapid growth only after the primary SASI mode
has saturated, the toroidal kinetic energy in both models with an $x$ sloshing
SASI mode grows exponentially from
the beginning. This illustrates the larger amount of numerical noise
experience by this mode given the grid geometry.

Using the transmitting boundary condition results in a lower amplitude
of the toroidal kinetic energy at early times, by a factor of several, relative to using
a reflecting boundary condition. Nonetheless, both types of boundary condition lead to very
similar results in the evolution of the primary mode. The lower level
of noise motivates the use of the transmitting boundary condition 
as the default for production runs.

\begin{figure}
\includegraphics*[width=\columnwidth]{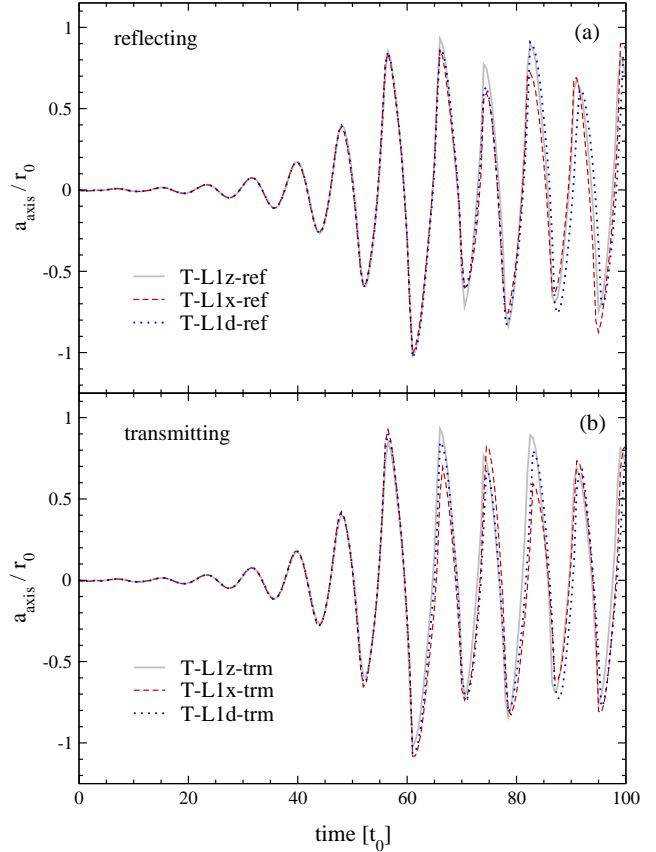}
\caption{Test of the isotropy of the code.  Curves show the real spherical harmonic coefficient
(eq.~[\ref{eq:a_i}]) corresponding to the axis along which a sloshing SASI mode 
is initially excited: $\hat z$ (solid gray),
$\hat x$ (dashed red), and a diagonal axis ($\hat z/\sqrt{2} + [\hat x + \hat y]/2$, dotted blue).
The top panel shows models that use a reflecting boundary condition in the $\theta$ direction
at the polar axis, and the bottom panel shows models that employ a transmitting boundary condition 
(\S\ref{s:axis_bnd}).}
\label{f:isotropy_test}
\end{figure}

\subsubsection{Symmetry along an arbitrary axis: isotropy}

The final test diagnoses the isotropy of the code, which we
assess by comparing the evolution of sloshing SASI modes
excited along different directions: (1) the $z$ axis, which
is the natural symmetry axis of the grid given the coordinate
system, (2) the $x$ axis, which while orthogonal to the $z$
axis is still aligned with a coordinate line on the equatorial plane,
and (3) a diagonal axis defined by
\begin{equation}
\hat d = \frac{1}{\sqrt{2}}\hat z + \frac{1}{2}\left( \hat x + \hat y\right).
\end{equation}
This direction lies along $\theta=\phi=45^\circ$, and hence
it is not aligned with any cartesian coordinate direction.

Figure~\ref{f:isotropy_test} shows the real $\ell=1$ spherical harmonic 
coefficients (eq.~[\ref{eq:a_i}]) along the corresponding sloshing axis for the last six
models of Table~\ref{t:test_models}. The evolution is identical in all cases
up to the point where the primary SASI mode saturates, demonstrating
the correctness of the implementation.

Models that use a reflecting boundary condition at the axis show
mutual agreement in the non-linear phase for a longer time relative to models
with a `transmitting' axis. This agreement extends particularly
to the maximum amplitude of the sloshing mode. In contrast, models
with `transmitting' boundary condition show closer agreement
in the oscillation phase at late times, but the amplitudes are not identical. 
The root-mean-square fluctuation of the real spherical harmonic coefficient,
computed between $t=75$ and $150t_0$ is shown in Table~\ref{t:test_models}.
All values are within a few percent of each other, and within a few
percent of the 2D value (0.544). 

\bibliographystyle{apj}
\bibliography{ccsne,apj-jour}

\label{lastpage}
\end{document}